\documentclass[twocolumn,twoside]{IEEEtran}

\usepackage{amsthm}
\usepackage{setspace}
\usepackage[margin=0.5in]{geometry}
\usepackage{booktabs}   
\usepackage{amsmath}
\usepackage{amssymb}
\usepackage[dvips]{graphicx}     
\usepackage{color}   
\usepackage[tight]{subfigure}
\usepackage{bm}
\usepackage{hyperref} 
\usepackage{color}

\newcommand{\mb}{\mathbf}

\title{Actively Calibrated Line Mountable Capacitive Voltage Transducer For Power Systems Applications} 
\author{Raffi Avo Sevlian and Ram Rajagopal
\thanks{R. Sevlian is with the Department of Electrical Engineering and the Stanford Sustainable Systems Lab, CA, 94305. Email: rsevlian@stanford.edu.}
\thanks{R. Rajagopal is with the Stanford Sustainable Systems Lab, Department of Civil and Environmental Engineering, Stanford University, CA, 94305. R. Rajagopal is supported by the Powell Foundation Fellowship. Email: ramr@stanford.edu.}
\thanks{This research was supported in part by the TomKat Center for Sustainable Energy, the Precourt Institute for Energy Efficiency, and the Thomas V. Jones Stanford Graduate Fellowship in Science \& Engineering.} 
}  

\begin{document}
\maketitle

\begin{abstract}  
A class of Actively Calibrated Line Mounted Capacitive Voltage Transducers (LMCVT) are introduced as a viable line mountable instrumentation option for deploying large numbers of voltage transducers onto the medium and high voltage systems.
Active Calibration is shown to reduce the error of line mounted voltage measurements by an order of magnitude from previously published techniques. 
The instrument physics and sensing method is presented and the performance is evaluated in a laboratory setting.
Finally, a roadmap to a fully deployable prototype is shown.
\end{abstract}     
 
\section{Introduction}     
  
Modern smart grid infrastructure envisions extensive deployment of renewable generation, electric vehicles, storage and many other technologies.
To enable these deployments, increased sensing and enhanced situational awareness is required.
In distribution systems, voltage and current are the main quantities which must be monitored and deploying technologies to measure these in a large scale can be prohibitively expensive.
Therefore, there is a need to develop inexpensive yet accurate sensor technology for the distribution system.
This work presents a new sensing technology which aims to dramatically reduce the cost of accurate voltage sensing for distribution system.

The state of the art in voltage measurement technology is either (1) expensive to deploy in large numbers or (2) suffers from errors making any practical use impossible.
Accurate voltage sensing in a substation setting relies on Voltage Transformers or Capacitively Coupled Voltage Transducer (CCVT)  \cite{ABB_CCVT} technology to step down the voltage.
Inexpensive line mounted voltage relies on (1) electrostatic field and (2) capacitively coupled measurements which do not reach the required accuracy of metering quality sensing.

This work proposes a solution to the limitations of line mounted voltage sensing.
The main contributions of this work are the following:
(1) A detailed physics model of Line Mounted Capacitive Voltage Transducers is presented with experimental validation.
(2) The active calibration methodology is detailed along with experiments showing its effectiveness of reducing the voltage magnitude errors.
(3) A research roadmap to enable a full deployable solution is given.

The paper is organized as follows.  
Section \ref{section-line-mounted-voltage-measurement-methods} reviews existing line mounted voltage measurement approaches. 
Section \ref{section-Physical-Model-of-Capacitive-Voltage-Transducer} presents the basic line mounted capacitive voltage transducer. 
Section \ref{section-Sensing-Methodology} introduces the active calibration concept and discusses some of it's properties.
Section \ref{subsection-device-prototype} describes the development of a practical prototype. 
Section \ref{section-Active-Calibration-Algorithms} presents in detail the signal processing methods required.
Section \ref{section-Active-Calibration-at-High-Votlage} presents experimental results proving the efficacy of the technology.
Finally, Section \ref{section-conclusion} addresses future challenges and opportunities for the proposed sensing technology. 

\section{Line Mounted Voltage Measurement Methods}
\label{section-line-mounted-voltage-measurement-methods} 
This section reviews the main classes of line mounted voltage measurement methods and identifies their benefits and limitations. 
\subsection{Electrostatic Field Measurement}
\label{subsection-Electrostatic-Field-Measurement}
The electric field generated by an energized power line is measured inside the device at one or more points at varying distances from the line.
The measured field is then used to reconstruct the line voltage based on a fixed mapping between the measured field and line voltage.
Much of the research on electric field based measurements has focused on the MEMs transducer themselves, \cite{Noras2010} \cite{Riehl2003}, \cite{Hsu1991} \cite{Davies1990}.
The fundamental limitation for the line mounted electrostatic sensors is the dependence on the physical arrangement of the conductor and the ground which dictates the observed electric field \cite{Bracken1976} \cite{Gerrard1998}, \cite{Miller1967}. 
This can change over time, but must be calibrated before sensor deployment typically in a laboratory setting.
When the physical environment changes over time, the inferred voltage level will deviate from the true value.

\subsection{Capacitive Voltage Measurement}  
\label{subsection-Capacitive-Voltage-Measurement}
The term capacitive voltage measurements enumerate a number of different configurations of traditional and non-traditional voltage measurements:
\begin{enumerate}
\item [1] \textit{Line/Ground Mounted} devices are electrically connected both to ground and the line.  
This method covers CCVTs which are installed in high voltage substations.
\item [2] \textit{Ground Mounted} devices are connected to ground and capacitively coupled to the energized line.
This method is most often associated with "Capacitively Coupled" measurement and is considered non-contact since it does not need electrical connection to the energized line.
The method has found uses in non-contact voltage instrumentation \cite{Noras2010}.
However, issues with multiple interfering power lines and large distance between the ground and the conductors have limited the accuracy. 
Some applications have been capable of estimating the power line phase arrangements \cite{Li2006}, even with large sensor errors.
\item [3] \textit{Line Mounted} devices are connected to the energized line and capacitively coupled to earth ground.
This method requires electrical connection to the energized line, but does not require grounded connection.
The paper focuses on this method since it is the only method that is line mountable, easy to deploy but also (as will be shown) allows for an \textit{active calibration} procedure enabling high accuracy.
\end{enumerate}      
  
There has been some attention paid to line mounted capacitively coupled voltage transducers for high voltage applications.
Two previously published patents: \cite{Smith1991} and \cite{Gunn2007} introduce 'body capacitive voltage measurement' similar to those shown here.
In \cite{Smith1991}, the authors state the use of a calibrated capacitor divider circuit for determining the voltage on the line.
The device is a doughnut shaped conductive material with an identical charge amplifier circuit presented in Section \ref{subsection-Charge-Transducer-Circuit}.
In \cite{Gunn2007}, the authors introduce a 'body capacitive' probe comprising of a fixed size sphere which hangs on the power line.
They present the sensing circuitry to measure the accumulated charge on the device as well.
Like \cite{Smith1991}, the calibration of the probe capacitance is done offline.  

In \cite{Yang2011} and \cite{Moghe2014}, the authors develop a similar understanding of the capacitive coupling and propose methods to track or mitigate changes in probe capacitance.
In \cite{Yang2011}, multiple conductors are used to mitigate the effect of nearby conductors.  
The results show a nominal voltage magnitude error of $1-12 \%$ with 1-5 minute averaging periods.
In \cite{Moghe2014} an algorithm is proposed to mitigate the nearby conductors and determine the height of the device from ground.
A parametric model relating the unknown height of the device from ground is used along with long time captures to estimate the height of the device and the probe capacitance.
Finally, the estimate of the probe capacitance is used to estimate the line voltage.

From these the prior and proposed LMCVT technology can be classified in the following classes.
\begin{itemize}
\item [1] \textit{Offline Calibration} estimates the voltage magnitude from a fixed mapping function which is determined beforehand in laboratory testing \cite{Smith1991}, \cite{Gunn2007}.
All electrostatic sensor based methods will fall into this category as well, since they rely on previously computed mapping function.
\item [2] \textit{Online Passive Calibration} tracks specific changes such as the height of the device or proximity to conductors by processing multiple passive sources \cite{Yang2011}, \cite{Moghe2014}. 
The changes are tracked in an online manner so that the mapping function changes over time, however the task is performed by processing passive measurements.
The commonality in both methods is that passive calibration requires (1) long captures (2) parametric models of disturbances and probe-to-ground capacitance.
\item [3] \textit{Online Active Calibration} uses active voltage injection onto the capacitive probe and then recover the perturbed value (depending on the probe to ground capacitance).
Careful signal design and signal processing techniques track in real time changes to the probe capacitance and estimate the line voltage.
\end{itemize}

The third method is what is proposed here to enable low cost high and medium voltage measurement technology.
Unlike passive techniques, this method does not require parametric models, and can compute the capacitance directly via pilot signaling mechanism.


\section{Physical Model of Line Mounted Capacitive Voltage Transducer}
\label{section-Physical-Model-of-Capacitive-Voltage-Transducer}

Various properties of a body capacitive probe are modeled through the first principles of the devices physical operation.
First the \textit{charge accumulation} of an ideal conductor held at the high voltage while leads to the floating capacitor is introduced.
Then, the \textit{effective capacitance} of the system is discussed as well as interference effects on nearby conductors.

\subsection{Ideal Body Capacitive Probe Model}
\label{subsection-Ideal-Probe-Model}
\begin{figure}[h]
\centering
\subfigure[][]{
\label{fig:CCED_free_space}
\includegraphics[scale=0.4]{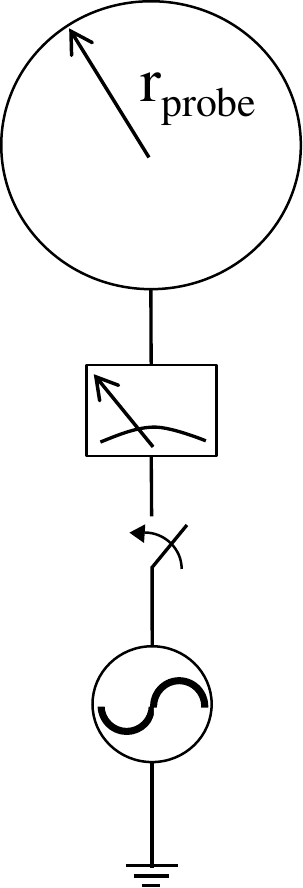}   
}
\hspace{-4.5mm}
\subfigure[][]{
\label{fig:CCED_with_proximity}
\includegraphics[scale=0.4]{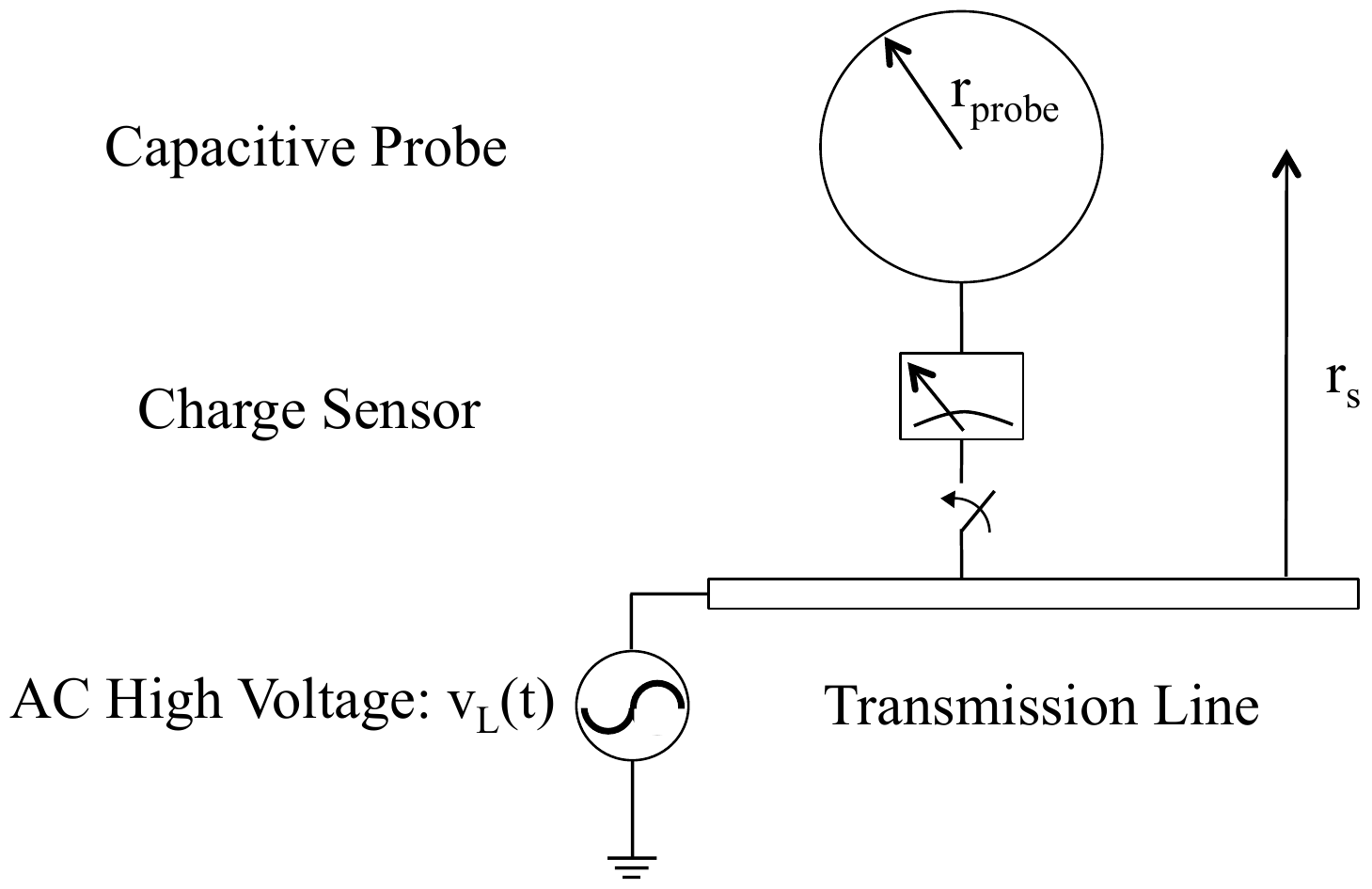}   
}
\caption{
\ref{fig:CCED_free_space} Ideal LMCVT consisting of a body capacitive probe connected to an ideal charge sensor connected to AC voltage source.
\ref{fig:CCED_with_proximity} Ideal LMCVT with proximity to nearby conductor.}
\end{figure}
Figure \ref{fig:CCED_free_space} introduces an ideal body capacitive probe.
The probe is a conducting sphere of radius $r_{probe}$.
Although it is impractical for a final device to be a round sphere, this model is used since it's capacitance is simple to compute.

Connected to the sphere is an ideal charge sensor which measures the charge that accumulates on the surface of the conductor.
The model assumes that the voltage source and the sensor take infinitely small volume compared to the conducting sphere, therefore producing no electric field of its own.
(the case of non-negligible voltage source is considered in \ref{subsection-Ideal-Probe-with-Power-Line}).
Furthermore, there is no voltage drop between the ideal charge sensor and the conducting sphere.

Assume at time $t \leq t_0$ the device is previously being uncharged and after $t > t_0$ the device is connected to the voltage source. 
Calculating the accumulated charge on the sphere necessary to maintaining a voltage of $V_{L}$ leads to $Q =  4 \pi \epsilon_{0} r_{probe} V_L$.
This is the basic definition of the probe capacitance to ground, where $C = 4 \pi \epsilon_{0} r_{probe}$.
The capacitance follows the standard relationship $Q = C V_{L}$ which leads to a very simple voltage transducer.
If the value of $C$ is known with certaintly, for example by building a probe with a spherical shape of known radius placed in free space, then measuring the waveform of $Q(t)$ gives us the waveform for $V_{L}(t)$.
\subsection{Ideal Probe with Power Line}   
\label{subsection-Ideal-Probe-with-Power-Line}

The proximity of the capacitive probe to the charging power line, or any other charged conductor will decrease the effective capacitance of the sphere.
Consider Figure \ref{fig:CCED_with_proximity} with the capacitive transducer but now attached between the device and the ideal voltage source is a power line.
The center of the conducting sphere is at a distance of $r_s$ from the center of the power line.

Assume a switch connecting the probe to the voltage source is switched on at some time $t_0$
At $t \leq t_0$ the cable is charged at $V_L$ but disconnected from the probe.
Given that the line is at $V_L$ there is a non-zero electric potential at various points in the system, $V(r, \phi, z)$.

Following the infinite length power line assumption, the voltage profile will have rotational symmetry as well as uniformity along the cable.
Therefore, only $V(r)$ needs to be considered.
At the moment $t > t_0$ only $Q = C (V_L - V(r_s) )$ amount of electrons need to accumulate on the conductor surface in order to bring the device to line voltage $V_L$.   
The voltage at surface of the conductor due to the accumulated charge is $V_L - V(r_s)$ while the contribution from the power line is $V(r_s)$ leading to both the power line and the device being charged to $V_L$.
This defines the effective capacitance $C_{p} \triangleq Q/V_L $ which holds regardless of power line proximity. 
Since $Q = C (V_L - V(r_s) )$, the probe capacitance is now: $C_p = C (1 - \alpha(r))$ where $\alpha(r_s) = V(r_s)/V_L$ is invariant to the actual voltage level and is computed from the system geometry.
Therefore, introducing the power line and voltage source only reduces the capacitance seen by the ideal charge sensor.
\subsection{Ideal Probe with Coupling Interference Source}
\label{subsection-Ideal-Probe-Coupling-Interference-Source}
\begin{figure}[h]
\subfigure[][]{
\label{fig:ideal_body_capacitive_probe_with_powerline_and_interference}
\includegraphics[scale=0.28]{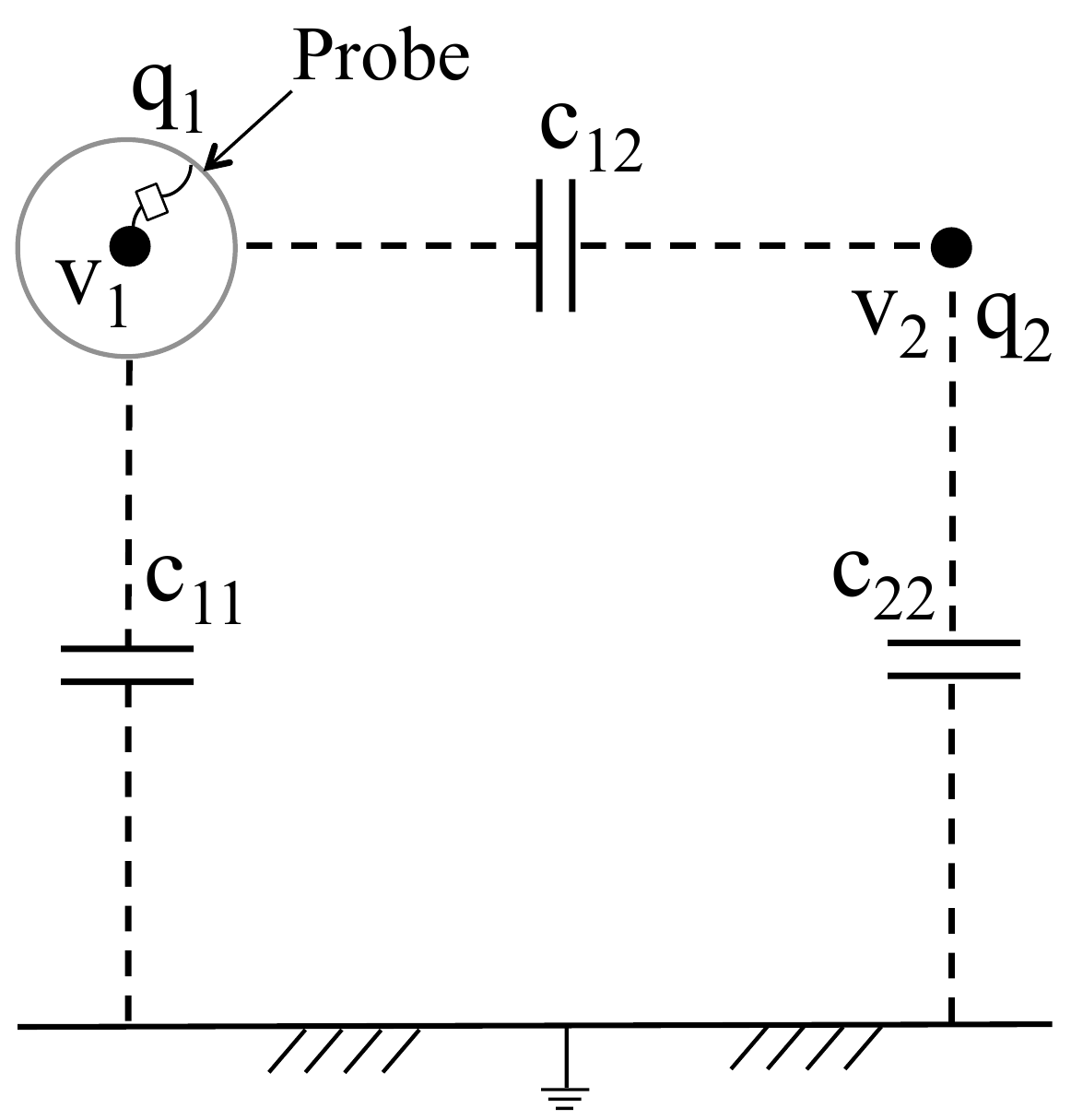}  
}
\subfigure[][]{  
\label{fig:system_circuit_equivalent_passive}
\includegraphics[scale=0.35]{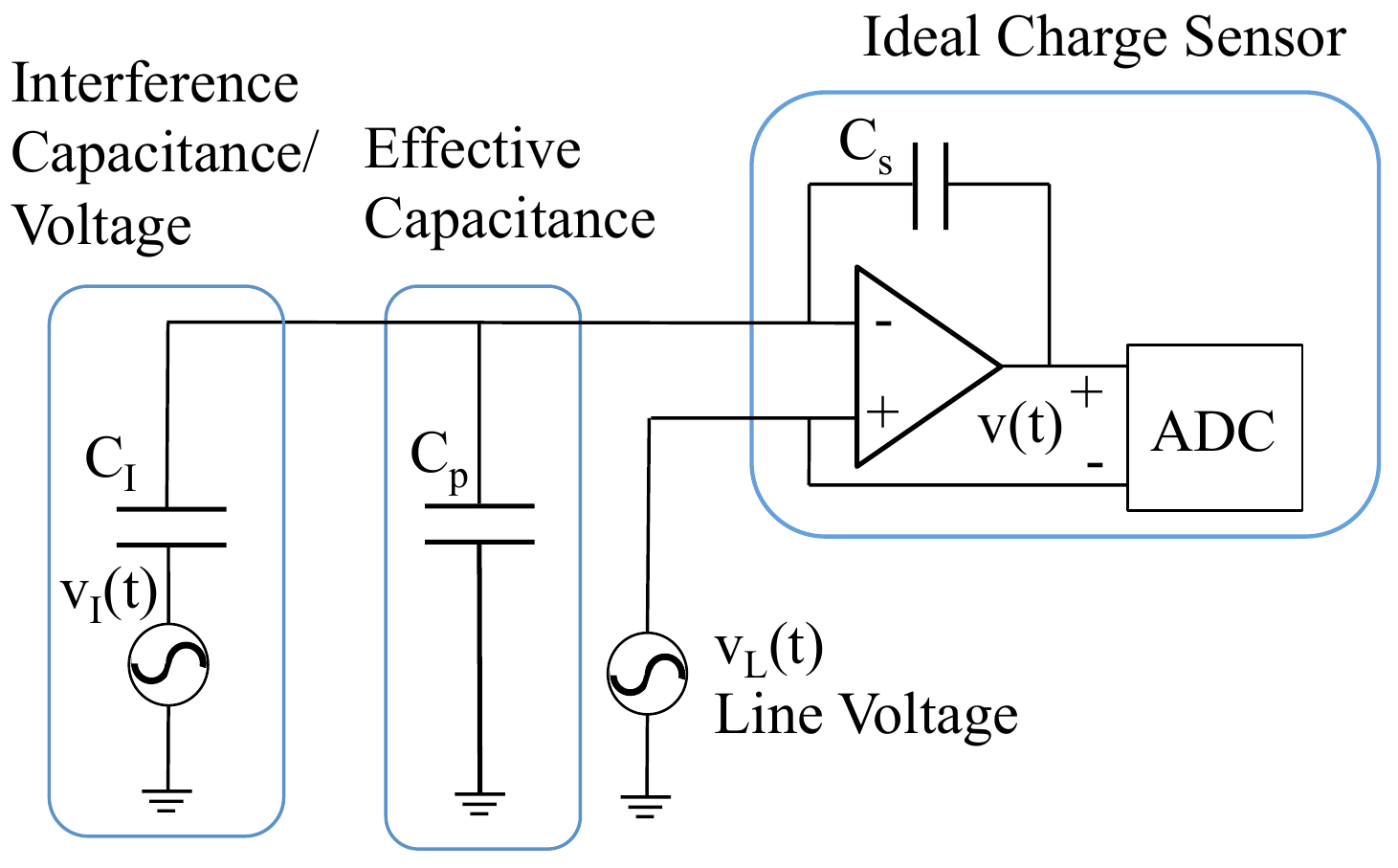}  
}
\caption{
\ref{fig:ideal_body_capacitive_probe_with_powerline_and_interference}
Body Capacitive probe as a multi conductor system. 
In a distribution system, the distance between conductors is small enough whereby $c_{11} + c_{12}$ is an order of magnitude larger than $c_{12}$.
\ref{fig:system_circuit_equivalent_passive}
Circuit diagram of passive LMCVT device with practical charge sensing capability, probe capacitance $C_p$ and interference capacitance $C_{I}$.
}   
\end{figure}
Coupled measurements between nearby conductors is an important effect that must be considered.
In a capacitive line sensor, the predominant form of interference is from crosstalk between the various lines. 
This work considers only a single interferer for now, but the results can be extended to multiple interfering power lines.
Figure \ref{fig:ideal_body_capacitive_probe_with_powerline_and_interference}, shows the ideal body capacitor connected to the high voltage power line.
This three conductor system, contains (1) the body capacitive probe $v_1$, (2) earth environment which is grounded and (3) the interferer $v_2$.

In a multi conductor system, a capacitance matrix describes the electrostatic geometry \cite{CapacitanceMatrix2014}, \cite{GriffithsEM1999}.
Given the arrangement in Figure \ref{fig:ideal_body_capacitive_probe_with_powerline_and_interference} the full electrostatic environment is described by:
\begin{align}
\left[ \begin{array}{c} q_1 \\ q_2 \end{array} \right] = 
\begin{bmatrix} c_{11} + c_{12}  & -c_{12}  \\ - c_{21} &  c_{22} + c_{12} \end{bmatrix} \left[ \begin{array}{c} v_1 \\ v_2 \end{array} \right] \label{eq:2by2cap_matrix}
\end{align}
Consider only the charge accumulation on probe $q_1$ since $q_2$ is of no interest.
Recall the effective capacitance $C_p$ is due to both the ground and the external environment.  
The cross term $c_{12}$ is the interference term $C_I$.    
This leads to
\begin{align}
v_1(t) &= (c_{11} + c_{12})  v_{L}(t) - c_{12} v_{2}(t) \\ 
          &= C_{p} v_{L}(t) - {C_I} v_{I}(t). \label{eq:full-cap-model}
\end{align}
\section{Sensing Methodology}
\label{section-Sensing-Methodology}
  
This section incorporates the physical model in Section \ref{section-Physical-Model-of-Capacitive-Voltage-Transducer} to develop a circuit representation of a passive LMCVT device.
Then the active calibration is introduced and it's practically implemented is shown.
  
\subsection{Circuit Model of Body Capacitive Sensor}   
\label{subsection-Charge-Transducer-Circuit}

The circuit equivalent of the body capacitive probe with both proximity effect and interference is shown in Figure \ref{fig:system_circuit_equivalent_passive}.
The ideal charge sensor used in Section \ref{subsection-Ideal-Probe-Model} which measures the charge induced can be implemented in practice by an op-amp with feedback capacitor $C_s$.

Assume the line voltage is $v_{L}(t) = V_L \cos(\omega t + \phi)$ and interference capacitance can be ignored, so $C_I = 0$
Calculate the output of op-amp (and high end of the differential ADC), $V^{ADC+}(j\omega)$.  
Since it is a non-inverting operational amplifier with feedback amplifier $Z_F$ and input impedance $Z_{IN}$ the output voltage is
\begin{align}
V^{ADC+}(j\omega) &= \left(1 + \frac{ Z_{F} } {  Z_{IN} }\right) V^{+}(j\omega) \\
			     &= \left(1 + \frac{C_p}{C_s} \right) V_{L}(j\omega).
\end{align}
Here, assume that the operational amplifier is ideal.  
In practice a low M $\Omega$ resistor is put in parallel with the capacitor to maintaining the leakage current of the device.
The low end of the differential ADC is $V^{ADC-}(j\omega) = V_L(j\omega)$.

So the differential voltage measured at the input of the analog input is:
 \begin{align}
 V(j\omega) &= V^{ADC+}(j\omega) - V^{ADC-}(j\omega) \nonumber \\ 
                   &=   \left(\frac{C_p}{C_s} \right) V_{L}(j\omega) \nonumber
 \end{align}
The addition of an interference source can be done via superposition principle. 
Shorting the AC voltage source and the body capacitor, leads to a negative feedback amplifier which results in the final form: $V(j\omega) = \frac{C_p}{C_s} V_{L}(j\omega)  - \frac{C_I}{C_s} V_{I}(j\omega)$ or in time domain:
\begin{align}
v(t) = \frac{C_p}{C_s} v_{L}(t)  - \frac{C_I}{C_s} v_{I}(t)\label{eq:ideal-cap-input-output-with-interference}
\end{align}
%
%
\subsection{Passively Calibrated LMCVT}
\label{subsection-Passively-Calibrated-LMCVTs}  
Given this circuit model, a passive LMCVT merely samples the signal in \eqref{eq:ideal-cap-input-output-with-interference}.
The digital signal $v[n]$ is then used to estimate the probe capacitance and the line voltage.
With offline calibration methods, $\hat{C}_p$ is assumed fixed and known.
The line magnitude is then recovered via: $\hat{V}_L = (C_s/\hat{C}_p) \hat{V}$, where $\hat{V}$ is the magnitude of the digital waveform $v[n]$.
Passive calibration techniques in \cite{Yang2011} and \cite{Moghe2014}, propose using modeling assumptions on $\hat{C}_p$ and $v(t)$ as well as long time captures to estimate $\hat{C}_p$ and finally  $\hat{V}_L$ as before.
\subsection{Actively Calibrated LMCVT}
\label{subsection-Actively-Calibrated-LMCVTs}
\begin{figure}[h]
\includegraphics[scale=0.4]{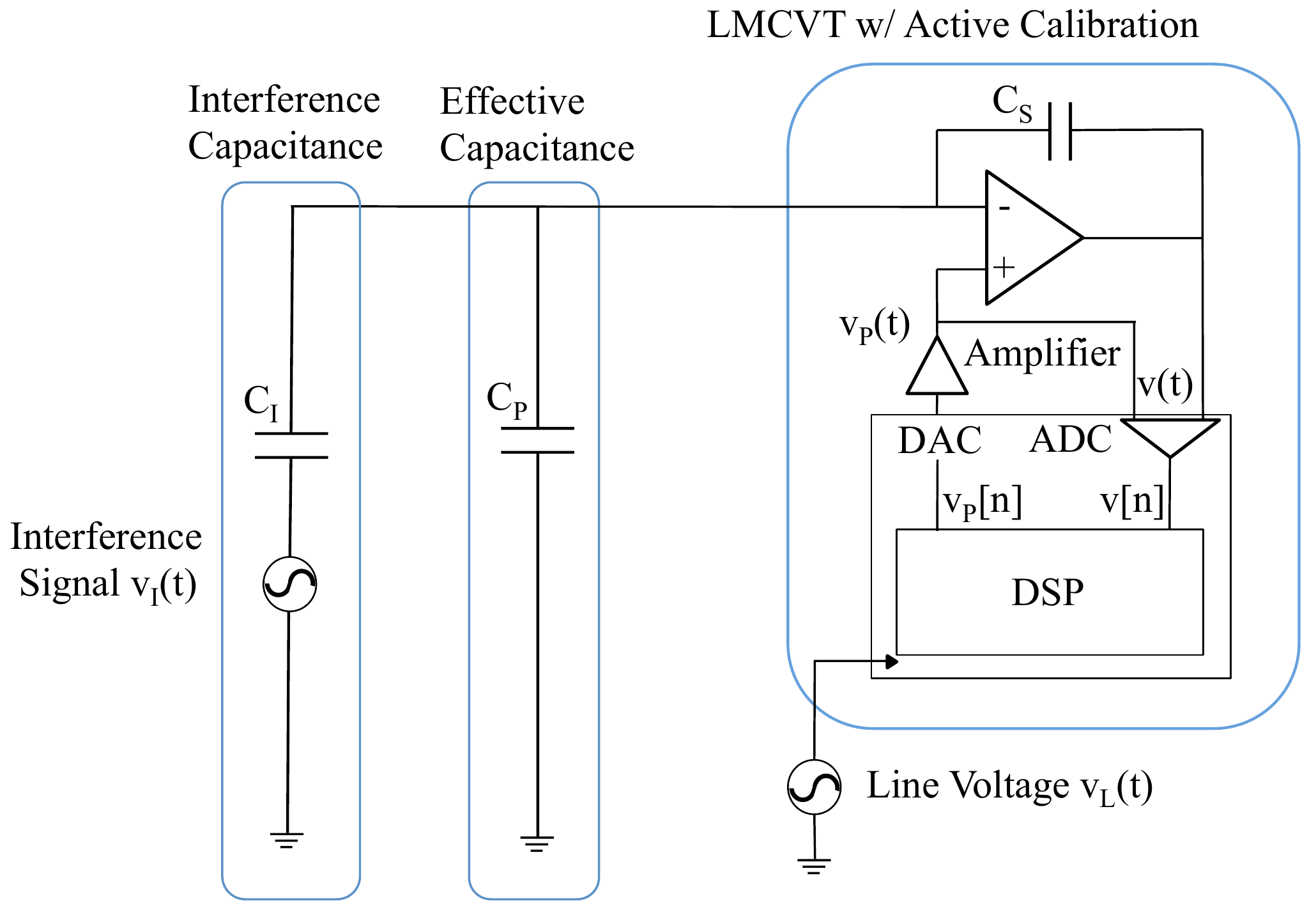}  
\caption{Equivalent circuit diagram representing basic voltage measurement.  
The body capacitor is shown as a shunt capacitance $C_p$ to ground. 
The interference source is capacitively coupled with value $C_I$.
The charge sensing device is built by a feedback amplifier with impedance $C_s$.}
\label{fig:system_circuit_equivalent_active}
\end{figure}  
Active calibration is an alternative method where \textit{one or more out of band pilot signals} close to 60 Hz be inserted between the line voltage and the charge sensing stage.
This can be performed with a general DSP platform, as in Figure \ref{fig:system_circuit_equivalent_active}.
In this situation, the device is energized at voltage level $v_{L}(t)$, and the DAC output of pilot signal is $v_{P}(t)$.
The non-inverting terminal voltage in an active calibration system is $v^{+}(t) = v_{L}(t) + v_{P}(t)$ vs. $v^{+}(t) = v_{L}(t)$ in the passive case.

From \eqref{eq:ideal-cap-input-output-with-interference}, omitting the interfering power lines, the input of the differential measurement is
\begin{align}
v(t) = \frac{C_p}{C_s} \left(v_{L}(t) + v_{P}(t) \right).
\end{align}
Since the line voltage and the pilot signal are at different frequencies, the received line voltage signal $(C_p/C_s) v_{L}(t)$ can be filtered, leaving only the term $(C_p/C_s) v_{P}(t)$.
The pilot signal is a known quantity, therefore, it is possible to use the known signal to recover $C_p$.
The practical implementation can be performed in an onboard DSP platform.
Active calibration eliminates the need of performing offline calibration for 'typical arrangements' of capacitive probes on transmission and distribution systems.

In theory, a small out of band pilot signal can estimate the probe capacitance at very high voltages.
Consider a 300 KV high voltage line, and an injected pilot signal of 10 V.
Given the typical body capacitance of $C_p = 20 pF$, if the maximum desired the input magnitude is $\pm 5$ V, then the feedback capacitor must be $600$ nF.  
Also, in this situation, the amplitude of the pilot in the ADC is 167 $\mu V$. 
This may be lower than the noise floor, but since the geometry in high voltage lines changes so infrequently averaging periods can be rather long.
In comparison, for a distribution line with nominal voltage of 10 kV, the minimum pilot voltage is 5 mV which will be close to but higher than the noise floor.

\subsection{Active Calibration and Line-Mounted Voltage Sensing}
\label{subsection-Capacitive-Coupling-and-Active-Calibration}
\begin{figure}[h]  
\label{fig:active_calibration_comparison_of_methods}
\subfigure[][]{
\includegraphics[scale=0.50]{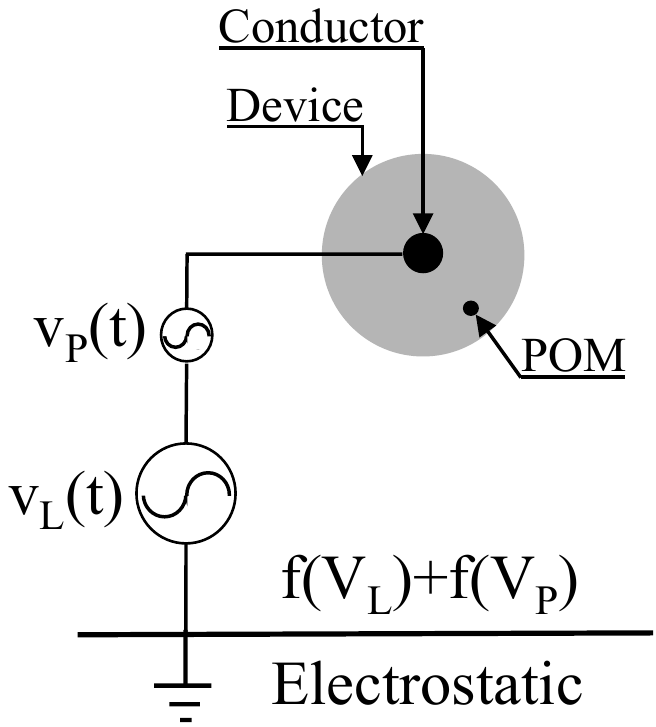} 
\label{fig:active_calibration_electrostatic_field_vs_capacitive_coupling_left}
}
\subfigure[][]{   
\includegraphics[scale=0.50]{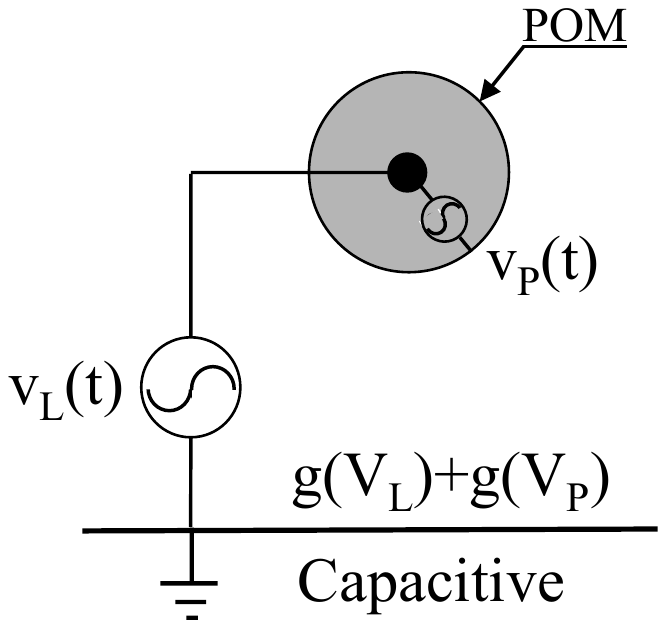}
\label{fig:active_calibration_electrostatic_field_vs_capacitive_coupling_right}
}
\caption{
Comparison of 'active calibration' mechanism in electrostatic field measurements (\ref{fig:active_calibration_electrostatic_field_vs_capacitive_coupling_left}) and capacitively coupled measurements (\ref{fig:active_calibration_electrostatic_field_vs_capacitive_coupling_right}).
Both methods include a line voltage source $v_{L}(t)$, perturbation voltage $v_{P}(t)$, device enclosure, point of measurement (P.O.M.), and contact conductor.
}
\end{figure}     
LMCVT technology is the only line mounted measurement technique that lends itself to active calibration.
To understand why, consider both the electrostatic field technique and capacitive coupling method shown in Figure \ref{fig:active_calibration_electrostatic_field_vs_capacitive_coupling_left}, \ref{fig:active_calibration_electrostatic_field_vs_capacitive_coupling_right}.  

In the case of electrostatic field measurement, the Point of Measurement (P.O.M) is a point inside the device. 
The measured value $E = f(V_L)$ depends on the physical configuration of the \textit{conductor and the remaining environment}.
Normally, some offline procedure is used to calibrate an inverse mapping $\hat{V}_{L} = f^{-1}(E)$, where $f^{-1}(\cdot)$ is fixed.
An active calibration must have a voltage perturbation on the entire conductor, to measure a perturbed output $f(V_{p})$ since the field depends on the changed conductor interacting with the environment.
In Figure \ref{fig:active_calibration_electrostatic_field_vs_capacitive_coupling_left}, the perturbation voltage must be placed on the entire conductor.

Alternatively, for capacitive coupling, the measured $Q = g(V_L)$ depends on physical configuration of the \textit{capacitive material and the remaining environment}.
In this case, active calibration needs to only to inject a voltage perturbation onto the conductive material to measure a perturbed output $g(V_{p})$ since the accumulated charge is caused by the conductive materials interaction with the environment.
This arrangement is feasible in a practical line mounted circuit.
In Figure \ref{fig:active_calibration_electrostatic_field_vs_capacitive_coupling_right}, the perturbation voltage can be placed between the conductor and the floating capacitor.
\section{Device Prototype}
\label{subsection-device-prototype}
\begin{figure}[h]  
\centering 
\subfigure[][]{
\label{fig:current_prototype}
\includegraphics[scale=0.255]{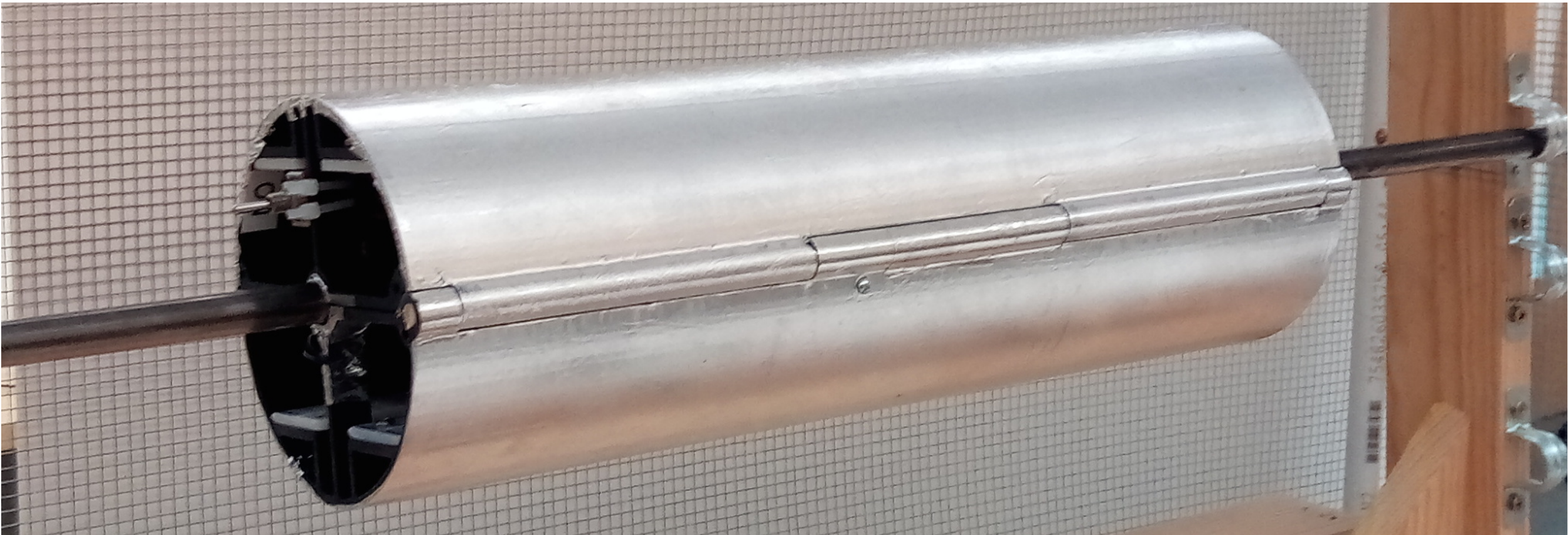} 
}
\subfigure[][]{
\label{fig:LMCVT_internal}
\includegraphics[scale=0.32]{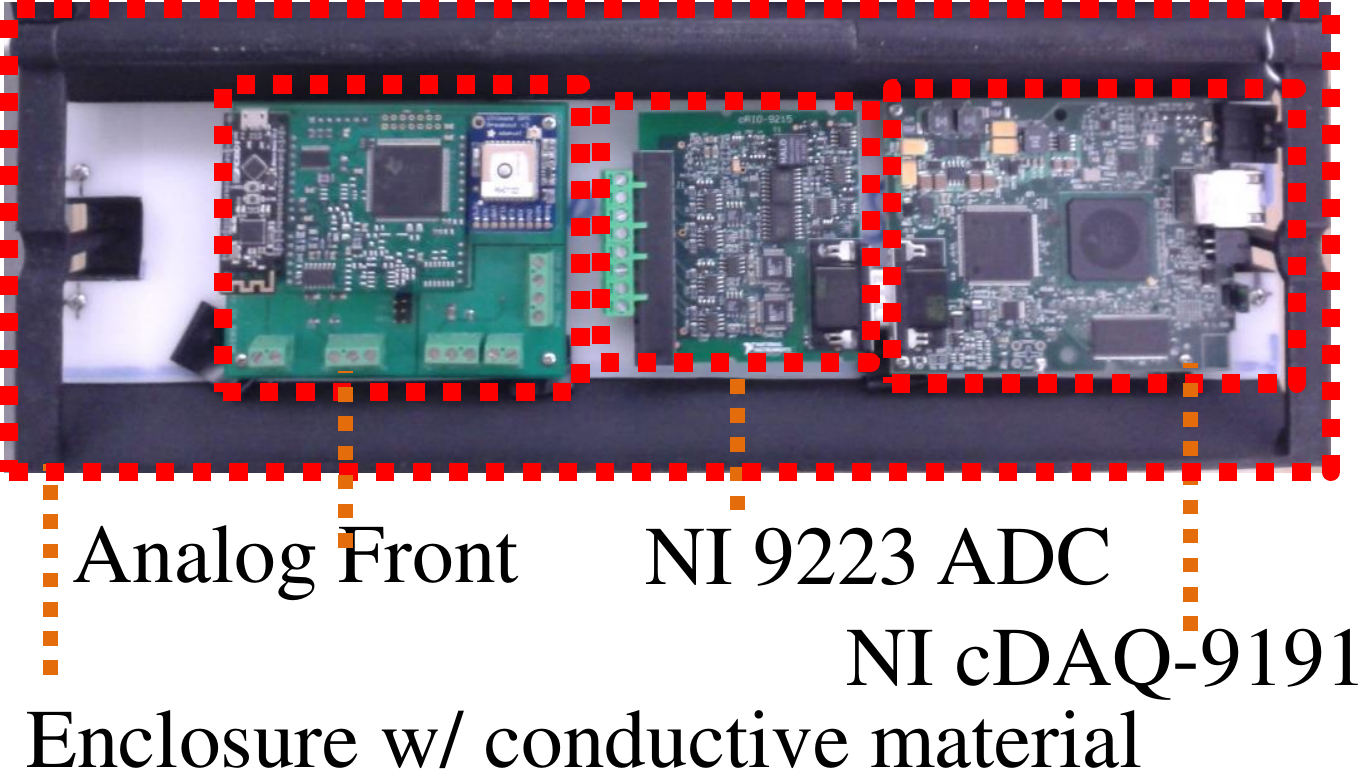} 
}
\subfigure[][]{
\label{fig:LMCVT_analog_front_end}
\includegraphics[scale=0.265]{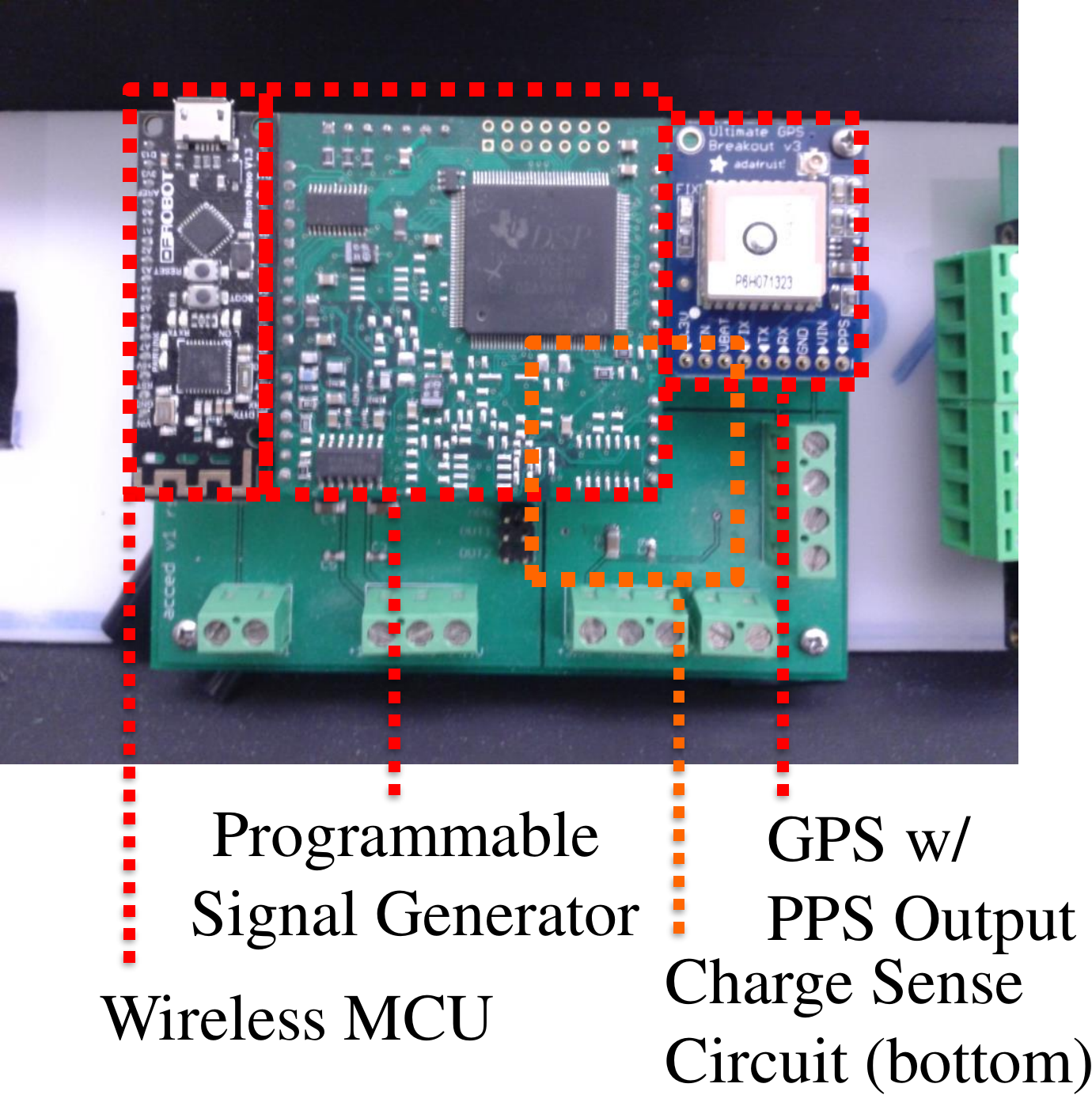} 
}
\caption { 
\ref{fig:current_prototype} Current prototype designed for further development and testing.
\ref{fig:LMCVT_internal} Internal electronics consists of analog front end board, data acquisition system, and wireless communication.
\ref{fig:LMCVT_analog_front_end} Analog front end consists of a wireless micro controller communicating with a programmable signal generator which injects the pilot signal into the system.  
Charge amplifier circuit to produce measurable output, and GPS PPS signal is outputted to the NI wireless ADC system for timing synchronization.
}  
\end{figure}
Figure \ref{fig:current_prototype} shows a line mountable prototype for evaluating the actively calibrated LMCVT technology.
The prototype includes a number of commercial off the shelf (COTS) and custom hardware components (shown in Figure \ref{fig:LMCVT_internal}):
\begin{enumerate}
\item Analog front end for pilot injection and signal recovery.
\item NI-9223 ADC and NI cDAQ-9191 will provide a 100 KS/s, 12 bit ADC and wireless streaming.
\item High density LiPo Batteries, which allow up to 5 Hrs of continuous capture time.
\end{enumerate}  
The Analog front end board consists of the following components (shown in Figure \ref{fig:LMCVT_analog_front_end}):
\begin{enumerate}
\item PCB mountable GPS with Pulse Per Second (PPS) output;
\item Low bias current op-amp and feedback capacitor for outputting the appropriate scaled down voltage signal; 
\item Frequency Devices SPPOSC-02 Series Programmable Oscillator \cite{FREQ_DEV_SPPOSC} which can be programmed to output pure and broadband pilot signals; 
\item Bluetooth enabled microcontroller to control programmable oscillator; 
\end{enumerate}
The device currently under development is shown in Figure \ref{fig:LMCVT_internal} and the analog front end in Figure \ref{fig:LMCVT_analog_front_end}.
The modular design allows testing of the technology and upgrades of each component.
The signal generator \cite{FREQ_DEV_SPPOSC} is capable of generating frequencies from 400 Hz to 100 kHz from an internal DSP.
%
\section{Active Calibration: Algorithms}   
\label{section-Active-Calibration-Algorithms}  
This section provides a high level architecture view of the tasks required for active calibration and detailed descriptions of each task block.
\vspace{-3mm}
\subsection{Signal Processing Architecture}
\label{subsection-Signal-Processing-Architecture}
\begin{figure}[h]
\label{fig:signal_processing_workflow}  
\includegraphics[scale=0.45]{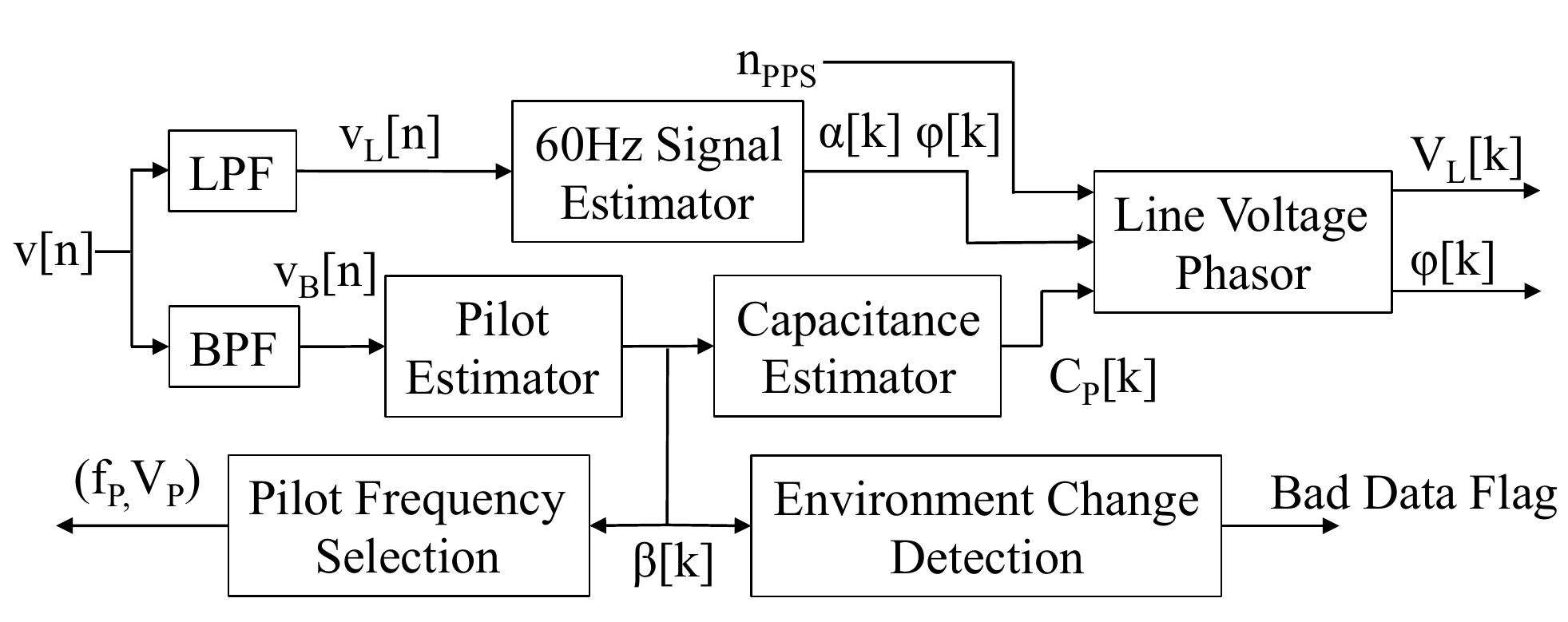} 
\caption{ 
Data flow for the LMCVT device for enabling PMU functionality.
Sampled data $v[n]$ is filtered: (1) with low pass filter (LPF) for line voltage phasor estimation (2) band pass filtered (BPF) for each pilot signal.
Variations on (NLLS) sinusoid estimator are used in extracting recovered pilot magnitude $\beta_1[k] \hdots \beta_M[k] $ and mains sinusoid phasor $\alpha[k]$, $\phi[k]$.
The pilot signals are used to (1) estimate the probe capacitance $\hat{C}_P[k]$ (2) invalidate signal output via bad data flag via environmental change point detection (3) adaptively set pilot magnitude and frequencies depending on frequency availability.
Finally, $\beta[k]$ and $\alpha[k]$ are used for outputting the final voltage phasor.
}
\label{fig:signal_processing_workflow}
\end{figure}
Figure \ref{fig:signal_processing_workflow} illustrates the general signal processing workflow devised to extract the line voltage from the received signal $v[n]$.
For clarity, Appendix Table \ref{tab:signal_processing_architecture} shows the different signals used in the architecture diagram.

\begin{enumerate}
\item [1] \textit{Line and Pilot Signal Estimation}: 
This step estimates $\alpha[k], \phi[k]$ the line magnitude/phase and $\beta[k] \hdots \beta_M[k]$, the pilot magnitude in a single cycle basis. 
Signal $\alpha[k]$ is used to recover line voltage, and $\beta_1[k] \hdots \beta_M[k]$ is used to recover the probe capacitance.
\item [2] \textit{Frequency Selector}:
Certain frequency bands can be periodically corrupted by interfering line sources.
To ensure clear frequency access, the recovered signal must be deemed reliable, if not a reliable pilot frequency must be determined.
This block controls the $M$ pilot frequencies and magnitudes: $f_{p, m}, V_{p,m}$.
\item [3] \textit{Probe Capacitance Estimator}:
Signals $\beta_1[k] \hdots \beta_M[k]$ are used to reconstruct the probe to ground capacitance $C_p$.
Ideally environmental changes are infrequent (on order of minutes or hours) so very accurate estimates can be made.
\item [4] \textit{Environmental Change Detector}:
The recovered pilot signal $\beta[k]$ is used to detect changes to the environment so that a portion of the system output is discarded.
After disturbances, the previous estimate data are discarded.
\end{enumerate}
 
In the following sections we elaborate each of the subcomponents.
Further work is needed to fully design optimal algorithms for these subcomponents.
\subsection{Sinusoid Estimation}
\label{subsection-Sinusoid-Estimation}
A  Non Linear Least Squares (NLLS) Estimation procedure is applied to both $v_L[n]$ and $v_B[n]$.
This reduces to a standard least squares signal estimator for the signal amplitude and phase, \cite{Stoica2005} and \cite{Stoica1989}.
The method is applied on a vector of length N, which is set by the sample rate and expected device output rate.
The nominal device output rate is $f_{frame} = 60 S/s$, which corresponds to a single cycle frame length.
The general least squares estimation problem is 
\begin{align}  
\{ \hat{\theta}, \hat{\omega} \}= \underset{\theta, \omega}{\arg\min} \sum^{N}_n \left( f_{\theta, \omega}[n] - v[n] \right)^2. \label{eq:least_square_sinusoid}
\end{align}
Given the sampled waveform $v[n]$ and the fitting function $f_{\theta, \omega}[n]$.
The fit function, $f_{\theta, \omega}[n]$, is a sum of $L$ sinusoids of various unknown frequency ($\omega$), amplitude ($\alpha_l$) and phase ($\theta$), given by
\begin{align}
f_{\theta, \omega}[n] &= \sum^{L}_{l=1} \alpha_l \sin ( w_l n + \phi_l ).     
\end{align}
Assuming that the number of sinusoids can be determined from prior knowledge or pre-processing step (FFT analysis), the problem can be split into two separate subproblems: (1) $\omega$ known/$\theta$ unknown (2) $\omega$, $\theta$ unknown. 
 
Both can be solved by the same computational steps:  if $\omega$ is known, $\hat{\theta}$ can be calculated by the following least square analysis.
Note the fit function can be rewritten as  
\begin{align}
f_{\theta, \omega}[n] &= \sum^{L}_{l=1} \underbrace{ \alpha_l \cos(\phi_l)}_{\gamma_l} \sin( w_l n ) + \underbrace{ \alpha_l \sin(\phi_l)}_{\eta_l} \sin( w_l n ) \nonumber
\end{align}
, where $\theta = [\gamma_1, \eta_1, \hdots, \gamma_L, \eta_L]^{T}$ is unknown.
  
The inference problem in \eqref{eq:least_square_sinusoid} is non-linear in $\alpha_l$ and $\phi_l$, but linear in $\gamma_l$, $\eta_l$, where $\gamma_l = \alpha_l \cos(\phi_l)$ and $\eta_l = \alpha_l \sin(\phi_l)$. 
The matrix $F(\omega, \theta) = \left[ f_{\theta, \omega}[0], \hdots, f_{\theta, \omega}[N] \right]^{T}$ is constructed, which is equivalent to $F(\omega, \theta) = M(\omega) \theta$, with the matrix 
\[ \hspace{-4mm} M(\omega) = \left[ \begin{array}{ccccc}
\cos(w_1 0) & \sin(w_1 0) & \hdots & \cos(w_L 0) & \sin(w_L 0) \\
\vdots              & \vdots             &            & \vdots              & \vdots \\
\cos(w_1 N) & \sin(w_1 N) & \hdots & \cos(w_L N) & \sin(w_L N)
 \end{array} \right] \] 
 and $\mathbf{v}$ $= \left[ v[0], \hdots, v[N] \right]^{T}$.
Assuming fixed frequency, the amplitude/phase component can be determined via linear least squares analysis: $\hat{\theta}(\omega) = \underset{\theta}{\arg\min}  \| \mathbf{v} - M(\omega) \theta \|^2$.
The least square solution is the following: $\hat{\theta}(\omega) =  (M(\omega)^{T}M(\omega))^{-1}M(\omega)^{T} \mb{v}$.
Finally given the $\gamma_l, \eta_l$ values, the sinusoid amplitude and phase compute by:
\begin{align}
\text{Amplitude}_l &= \sqrt{ \gamma_l^2 + \eta_l^2 } \label{eq:sinusoid_mag_estimate}  \\
\text{Phase}_l &= \tan^{ - 1}( \gamma_l / \eta_l ). 
\end{align}       
This formulation allows us to separate the estimation of the multiple sinusoid parameters from the estimation of the frequency of the line/pilot/harmonics.
Since the minimization in \eqref{eq:least_square_sinusoid} depends on fixed pilot/harmonic frequencies $\omega$, a second outer minimization can be performed to track each frequency.
In \cite{Stoica1989}, the authors present a number of techniques to quickly estimate the frequency of various harmonics.
This level of computation is required for tracking the pilot signal when it is surrounded by multiple harmonics.
This has practical reason also due to ADC clock drift: actively tracking frequency improves the estimate output.
\subsection{Capacitance Estimation}
\label{subsubsection-Capacitance-Estimation}

Assuming non-interfering pilot frequency, and no-environmental changes, the capacitance of the system is fixed in moderate timescale (minutes-hours)
Depending on whether single or multiple frequencies are used in the estimation, different estimation procedures can be deployed.
Here only a single pilot mechanism is discussd.

Assume a constant $C_p(t) = \overline{C}_p$ over a short time horizon.
In single pilot mode, the injected pilot signal is $v_{P}(t) = V_P cos( 2 \pi f_1 t + \phi_1 )$.
From \eqref{eq:ideal-cap-input-output-with-interference}, the recovered signal can be represented with the following linear equation: 
\begin{align}
\beta[k] = \frac{V_P}{C_s} \overline{C}_p + w[k]
\end{align}
for processing interval $k = \{1, \hdots, K \}$.
Here, $w[k]$ is additive noise with noise bandwidth coming from the bandpass filter in Figure \ref{fig:signal_processing_workflow}.
Over a long timeframe, where the probe capacitance is assumed fixed, and no environmental changes are detected.
  
The least square estimate of the probe capacitance is then:
\begin{align}
\hat{C}_P = \frac{C_s}{V_P} \left( \frac{1}{K} \sum_{0}^{K} \beta[k] \right) \label{eq:capacitance_estimation}. 
\end{align}
A similar least square method can be devised for multi-frequency pilot scheme, but is not explored here.

\subsection{Line Voltage Estimation}   
\label{subsubsection-line-voltage-estimation}
     
Finally given the estimate of the probe capacitance, $\hat{C}_P[k]$, the line voltage can be estimated from
\begin{align}
\hat{V}_{L}[k] = \frac{C_s}{\hat{C}_P[k]} \alpha[k]. 
\end{align}
Assuming that $\hat{C}_p[k] = \hat{C}_p, \forall k$.
The estimation interval of $\alpha[k]$ is set by the measurement output rate of the device.
%
%
%
%
%
\section{Active Calibration: Experiments}
\label{section-Active-Calibration-at-High-Votlage}
%

\subsection{Experiment Setup}
\label{subsection-experiment-setup}
The wall outlet voltage is connected to a variable transformer (variac) then to a step up transformer so that various test voltages can be generated (see appendix for further details).
The maximum achievable voltage under this technique is 1.28 kV.
This signal has a significant amount harmonic distortion.
However, achieving high accuracy under this condition lends confidence of the pilot mechanism working in an actual distribution and transmission lines where out of band noise is common.

In the following sections, tests at various voltages are performed, indicated as $\{T_0, T_1, \hdots T_9, T_{10} \}$.
This corresponds to variac positions of $\{ 0\%,~10,~\hdots 90 \%,~100 \% \}$ of the maximum voltage output.
The multimeter ground truth values are $V_{L, t}$, for $t = \{0, \hdots, 10 \}$, and given in Table \ref{tab:voltage_estimation_table}, Column 1.
%
\subsection{Received Signal}
\label{subsubsection-generating-line-voltage}

\begin{figure}[h]
\hspace{-6mm}
\subfigure[][]{
\label{fig:receive_adc_ts}
\includegraphics[scale=0.26]{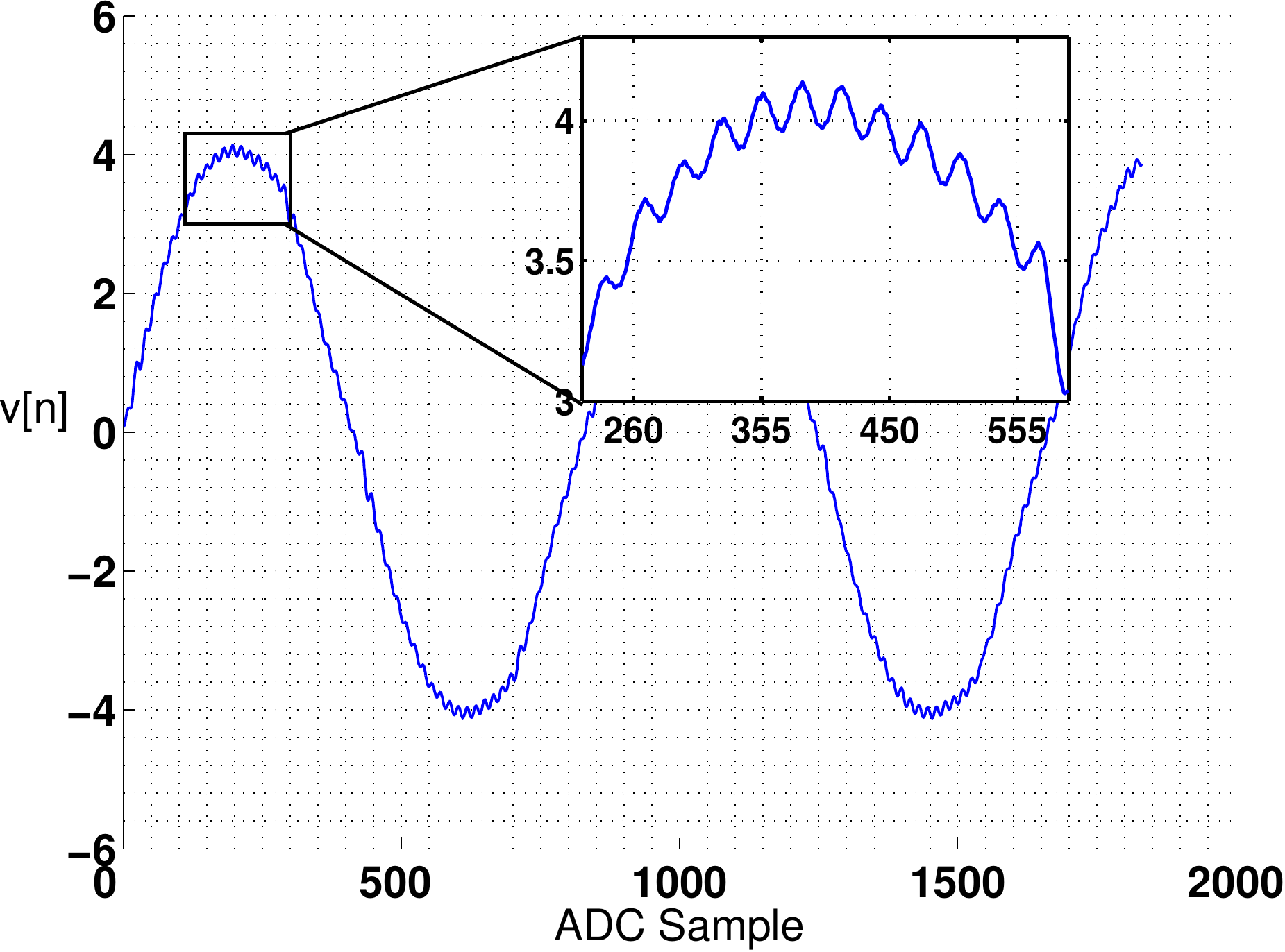} 
}
\hspace{-6mm}
\subfigure[][]{
\label{fig:receive_adc_bpf_fft}
\includegraphics[scale=0.26]{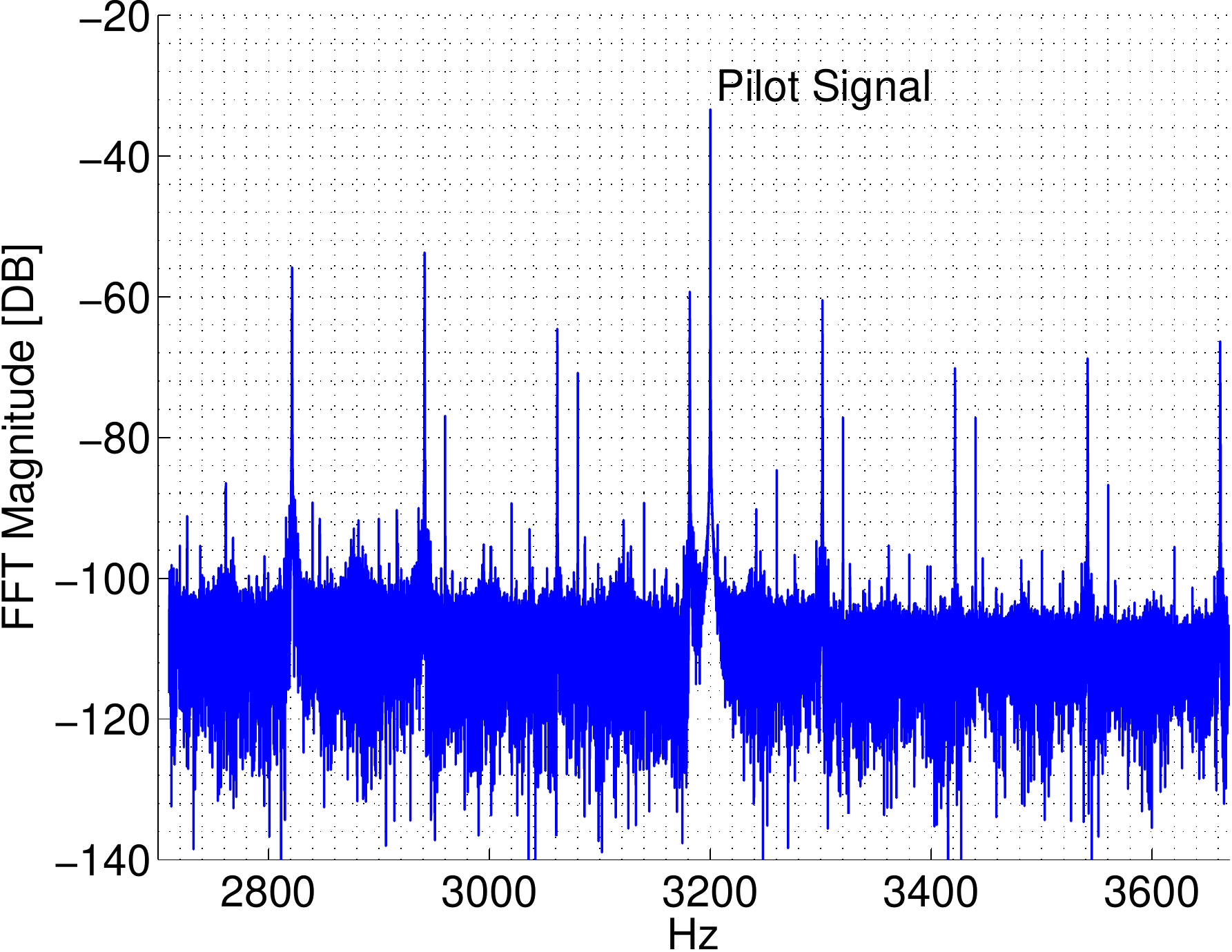} 
}
\caption{ 
\ref{fig:receive_adc_ts} Shows a typical captured time series from the LMCVT system.   
\ref{fig:receive_adc_bpf_fft} Shows the frequency region around the 3.2 kHz pilot signal recovered on $v(t)$.
Note prevalence of multiple harmonics of the main 60 Hz signal at amplitudes close to the pilot amplitude.
}
\label{fig:receive_signal_ts_fft}
\end{figure}

Figure \ref{fig:receive_signal_ts_fft}, shows the typical voltage waveform $v[n]$ in sample and frequency domain.
It is clear that the injected pilot signal is observed in $v[n]$, as shown in Figure \ref{fig:receive_adc_ts}.
However, looking in the frequency domain, Figure \ref{fig:receive_adc_bpf_fft}, the signal is close in magnitude with respect to the various harmonics of the mains voltage. 
These harmonics can be caused by (1) harmonic distortion from the ADC (2) harmonics of the 60 Hz signal that is normally present in the system.  
Figure \ref{fig:receive_signal_ts_fft} indicates that care must be taken in extracting the pilot amplitude.
\vspace{-4mm}
\subsection{Line and Pilot Signal Estimation}
\begin{figure}[h]
\hspace{-4mm}
\subfigure[][]{     
\label{fig:60Hz_amplitude_cage_nocage}
\includegraphics[scale=0.25]{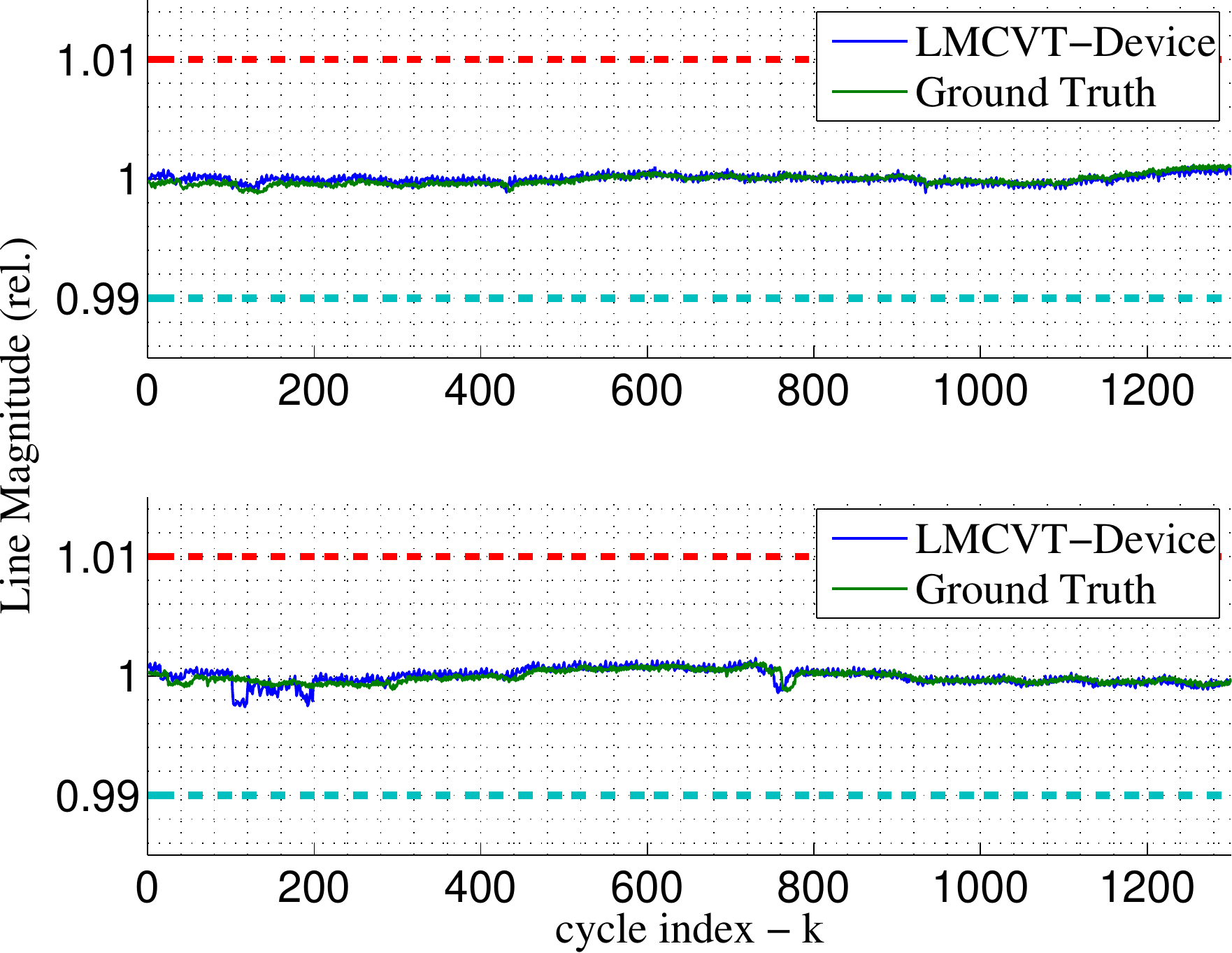} 
}   
\subfigure[][]{   
\label{fig:60Hz_amplitude_boxplot}   
\includegraphics[scale=0.26]{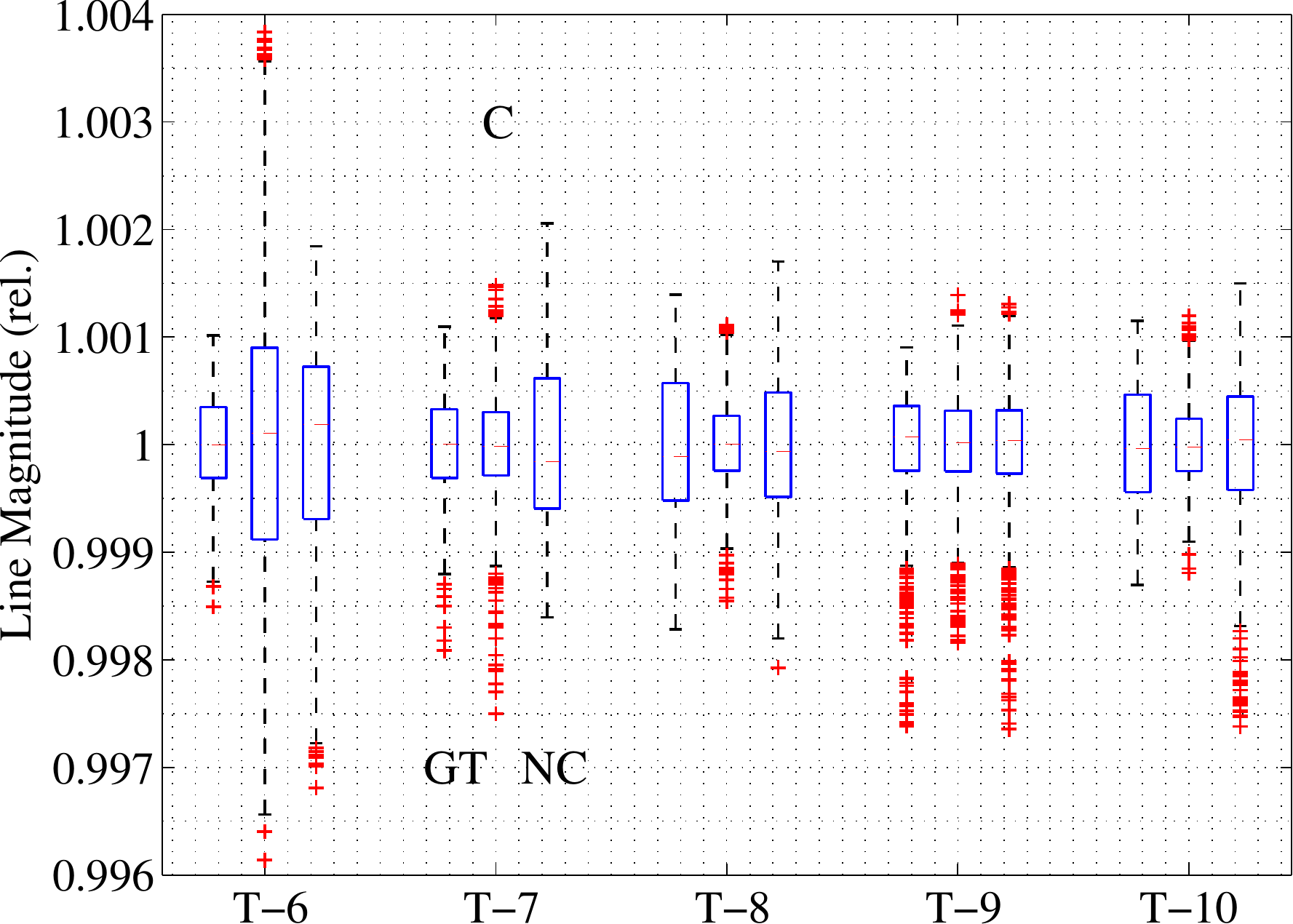} 
}  
\caption{
\ref{fig:60Hz_amplitude_cage_nocage}  
Normalized value of $\alpha_k$ for the LMCVT device and COTS resistor divider ground truth in restricted faraday cage setup (top subplot) and open air experiment (bottom subplot).
Time series show stability in the estimated signal amplitude.
\ref{fig:60Hz_amplitude_boxplot} 
Box plot for test $T_{7}, \hdots, T_{10}$.   
For each test, single cycle amplitudes shown in controlled, faraday cage (left) and outside, free floating (right) environment.
}
\end{figure}
The captured waveform $v[n]$ is low pass filtered to remove any high frequency components.
The NLLS procedure is applied on $v_{L}[n]$ under various test voltages.
Figure \ref{fig:60Hz_amplitude_cage_nocage} shows the recovered amplitude time series in a single cycle capture window.
In the case of tracking the line frequency, since the line magnitude is orders of magnitude larger than any harmonics, $v[n]$ can be low pass filtered and tracked with a single sinusoid $M=1$ model.
Note that the pilot magnitude, $\beta[k]$ is estimated identically as $\alpha[k]$, using \eqref{eq:sinusoid_mag_estimate}.

Figure \ref{fig:60Hz_amplitude_boxplot} show box plots of the ground truth (GT), controlled (C) and free space (NC) relative voltage magnitude captures.
The experiments indicate that the received voltage magnitude are comparable in variation.
This is very important since it indicated that under nominal conditions of fixed $C_p$, the variation between a traditional grounded instrument (GT), fixed floating capacitor (C), and open air float in capacitor (NC) are nearly identical.
As shown in section \ref{subsubsection-line-voltage-estimation}, the largest source of error is the estimation error of the probe capacitance, and not the variation of the line signal.
This makes sense since, the maximum variation of the line voltage $\alpha[k]$ is well within the $\pm 1\%$ limits for full capture variation.
\vspace{-4mm}
\subsection{Capacitance Estimation}
\label{subsubsection-Capacitance-Estimation}
\begin{figure}[h]     
\hspace{-5mm}   
\centering
\subfigure[][]{  
\label{fig:capacitance_estimate_single_pilot}
\includegraphics[scale=0.25]{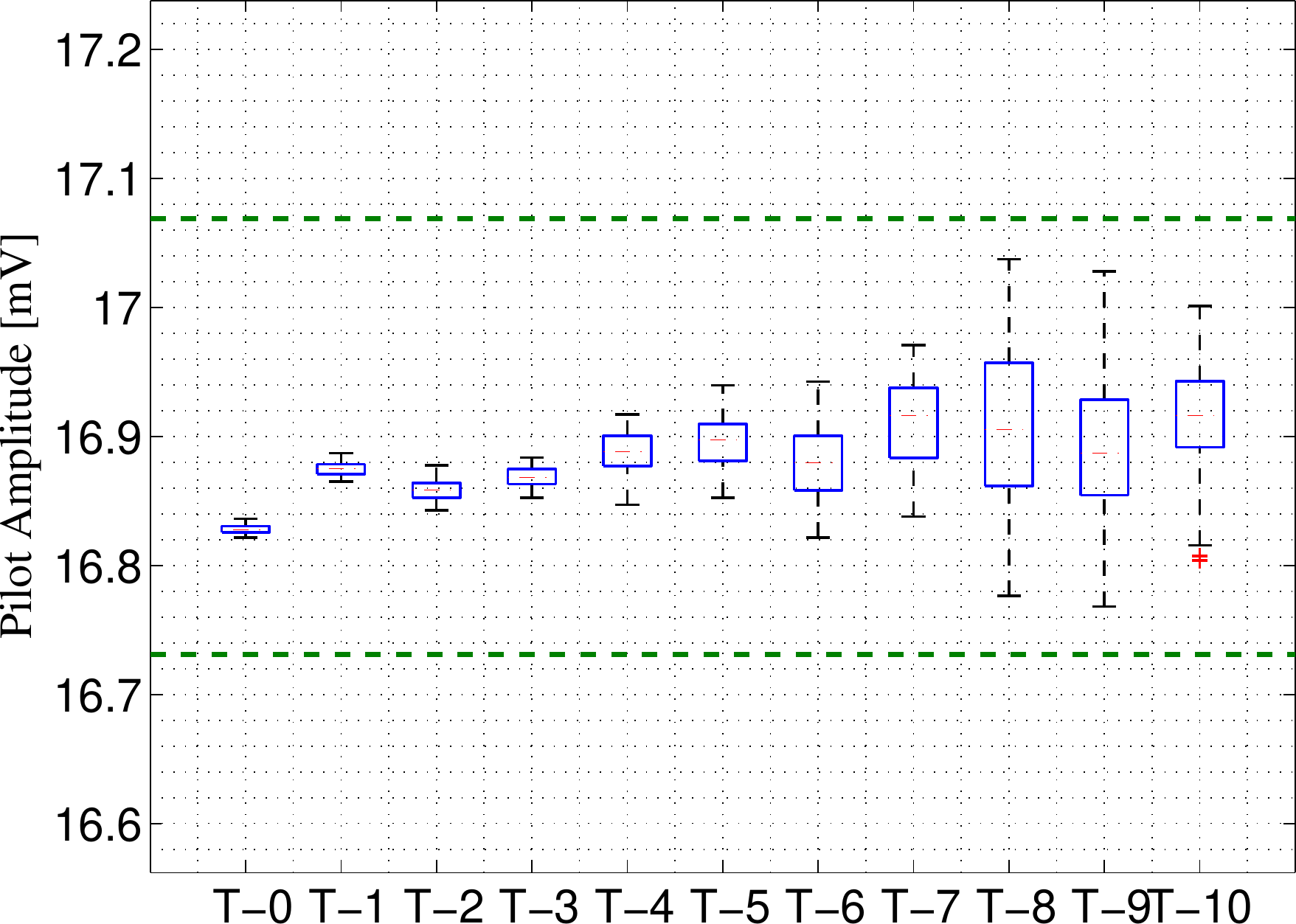} 
}
\hspace{-2mm}     
\subfigure[][]{  
\label{fig:capacitance_estimate_mean_and_cp_true}
\includegraphics[width=0.45\linewidth,height=0.35\linewidth]{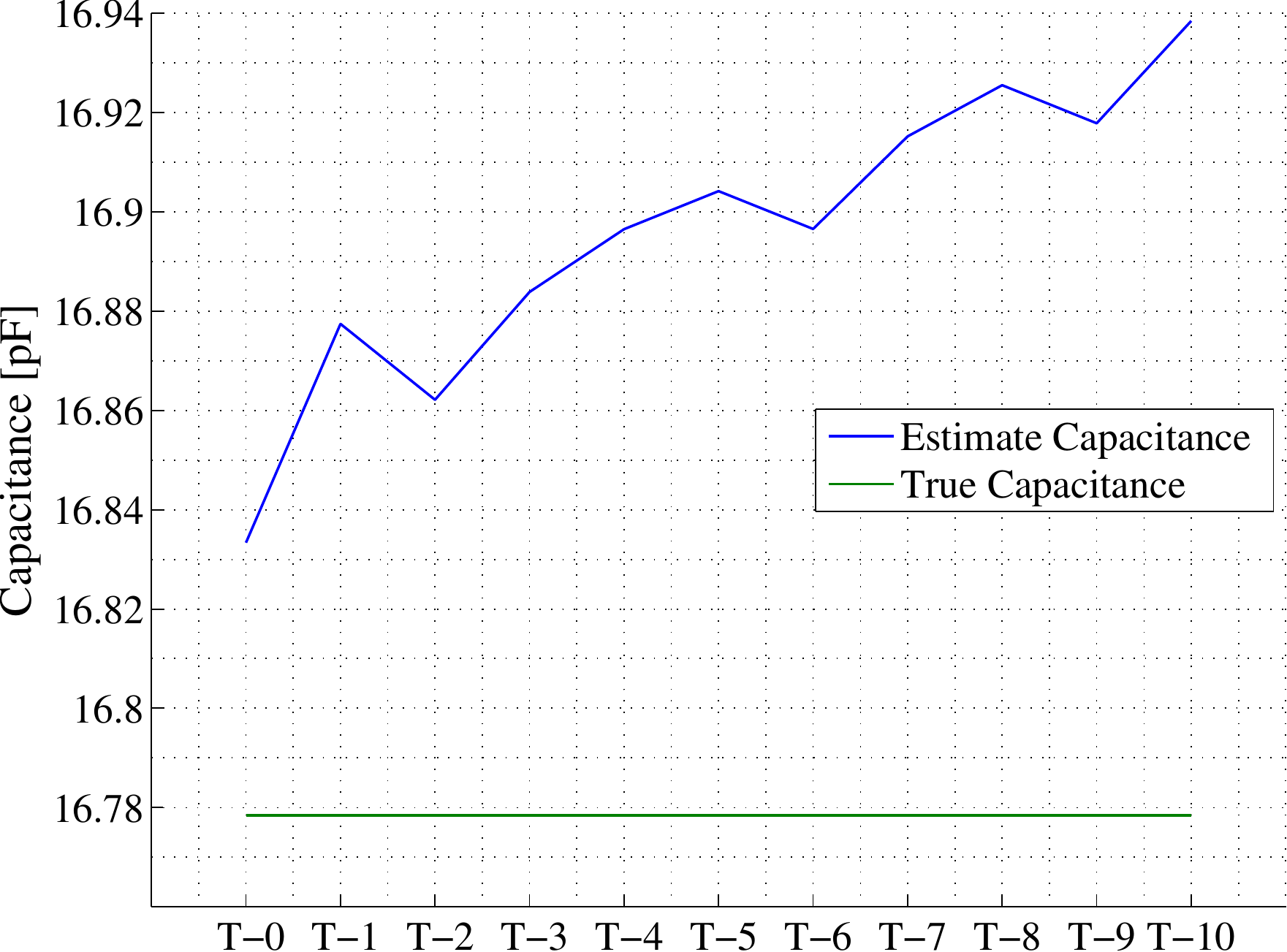} 
}  
\caption{ 
\ref{fig:capacitance_estimate_single_pilot} 
Recovered pilot magnitude $\beta[k]$ for K = 60 samples/second output rate.
Note that pilot signal is fairly stable over the range of line voltages under each test.
\ref{fig:capacitance_estimate_mean_and_cp_true} 
Mean pilot amplitude under each test scenario is used to estimate the probe capacitance.
}
\end{figure}         
Figure \ref{fig:capacitance_estimate_mean_and_cp_true} shows the estimate of $\hat{C}_p$ under each experiment. 
Although the signals look very close to each other, an upward bias is evident (blue).
A full scale estimate of the probe capacitance, $C^{\star}_p$, is shown as well (horizontal-green).
\begin{align}
\bar{\alpha_{t}} &= \kappa V_{L, t} + e_t, \label{eq:regression_model} \\
C^{\star}_p       &= C_s~\hat{\kappa} \label{eq-calc-true-cap} 
\end{align}
This is estimated as from the experimental data by solving the following regression equation solving for $\kappa$ in \eqref{eq:regression_model}.
Then \eqref{eq-calc-true-cap} calculates the true capacitance.
Here, $V_{L, t}$ is the mean voltage recorded on the digital multimeter ground truth and $\bar{\alpha_{t}} = \frac{1}{K} \sum_k \alpha_t[k]$.
Each variac test is an independent observation of \eqref{eq:regression_model}.

The error of a single cycle (K=1) estimate, using \eqref{eq:capacitance_estimation}, is the following: $\sigma(\hat{C}_p) = \left( \frac{C_s}{V_P} \right) \sigma_{W}$.
For the 1.2 kV test environment, $C_s =  10 \text{nF}$ and $V_P = 10 V$, however additive variance is extremely small.
To see why, the total additive noise on the signal is $3\sigma = 2 mV$ and since the signal is bandpass filtered, only a small portion of that initial noise will remain.
Typically, $\frac{ \sigma(\hat{C}_p) }{\hat{C}_p} \approx 3.8 \times 10^{-4}$.
So, the variation of the single cycle case, under moderate averaging intervals is fairly low.
A more important source of error is bias in the estimate, as discussed in Section \ref{subsubsection-line-voltage-estimation}, leads to most of the error.
\subsection{Line Voltage Estimation Results}     
\label{subsubsection-line-voltage-estimation}

\begin{table*}[!htb]
\centering     
\caption{Voltage Estimation Experiments.}   
\label{tab:voltage_estimation_table}   
\begin{tabular}{@{}lllllllll@{}}   
\toprule        
                  & & \multicolumn{3}{c}{$\hat{C}_p$ estimation via pilot}      & & \multicolumn{3}{c}{$C^{\star}_p$ full scale estimate}  \\ 
 $V_L$      & & $ \hat{V}_L$: mean/(range) [V] & & error : mean/sd/(range) [\%] & &  $ \hat{V}_L$: mean/(range) [V] & &  error : mean/sd/(range) [\%]   \\  
\cmidrule{1-1}    \cmidrule{3-3} \cmidrule{5-5} \cmidrule{7-7} \cmidrule{9-9}
  125.2    &  &  122.5 / (122.0   122.7)     & &   2.19 / 0.0638 / (2.0 2.5)             & &  123 (122 123)       & & ~1.61 / 0.0682 / (1.45 1.99)       		        \\   
  250.5    & &  250.1 / (249.6   250.5)      & &   0.19 / 0.0749 / (0.03 0.37)          & &   251 (250 251)      & & -0.30 / 0.0748 / (-0.46 -0.12)       	                \\   
  383.6    & &  384.0 / (383.3   384.3)      & &   0.19 / 0.0483 / (0.03 0.37)         & &   386 (385 386)      & &  -0.72 / 0.0462 / (-0.80 -0.53)                      \\   
  499.2    & &  502.6 / (501.7   503.1)      & &   0.68 / 0.0467 / (-0.07 0.49)        & &  506 (505 506)       & & -1.39 / 0.0420 / (-1.48 -1.20)                       \\   
  631.4    & & 629.3 / (626.9   629.8)       & &   0.68 / 0.0457 / (-0.07 0.49)        & &  634 (631 634)       & & -0.40 / 0.0408 / (-0.49 -0.03)                       \\   
  763.6    & & 752.8 / (749.9   753.3)       & &   1.42 / 0.0406 / (1.35 1.79)         & &  758 (755 758)        & &  ~0.73 / 0.0435 / (0.65 1.10)                         \\     
  890.9    & & 881.0 / (879.6   882.1)       & &   1.11 / 0.0533 / (0.99 1.27)         & &  888 (886 889)        & &  ~0.30 / 0.0513 / (0.18 0.46)                          \\   
  1019.6  & & 1011.5 / (1006.8  1013.0)  & &   0.80 / 0.0716 / (0.64 1.25)         & &  1020 (1015 1021)  & &  -0.06 / 0.0996 / (-0.22 0.38)                       \\   
 1146.9   & & 1136.3 / (1130.8  1144.6)   & &   0.92 / 0.0742 / (0.2   1.39)         & &  1145 (1140 1154)   & &  ~0.10 / 0.0751 / (-0.62 0.58)                        \\   
 1281.2   & & 1270.1 / (1266.5  1272.1)  & &   0.87 / 0.0696 / (0.71 1.14)         & &  1282 (1278 1284)  & &  -0.07 / 0.0618 / (-0.23 0.20)                       \\   
               & &                                         & &                                      & &                               & &                                                               \\   
               & &                                         & &  0.71 / 0.0589 / (0.52 1.01)           & &                               & &  -0.02 / 0.0593 /(-0.20   0.28)                      \\   
\bottomrule 
\end{tabular}
\end{table*}
       
Table \ref{tab:voltage_estimation_table}, shows the performance attained in the experiment where the the line voltage magnitude is estimated via active calibration. 
Table \ref{tab:voltage_estimation_table}, column 1, indicates the ground truth voltage which is measured by a digital multimeter measurement of the line signal.
Columns 2 and 3 indicate the results of the pilot based estimation. 
The output of the device is a single cycle estimate of the line voltage $V_L[k]$.
From this the single cycle relative errors: $e[k] = \frac{V_{L} - V_L[k]}{V_{L}}$ for each test, can be computed.
Therefore, following terms are reported on a single cycle basis:   
\begin{itemize}
\item [1] error mean $\mu(e) = \frac{1}{K} \sum_k e[k]$, 
\item [2] standard deviation $\sigma(e) = \sqrt{ \frac{1}{K} \sum_k (e[k] - \mu(e))^2 }$ 
\item [3] range: $r(e) = ( \min_k e[k], \max_k e[k] )$ 
\end{itemize}  
The results indicate that the current prototype is able to reach metering quality, in half of the test cases.
The mean relative error over all the tests is $0.72 \%$. 
This is because of the overall overestimate of the probe capacitance in each test as shown in Figure \ref{fig:capacitance_estimate_mean_and_cp_true} indicating a positive bias.
However, this may be removed by a multi-frequency pilot mechanism or higher frequency pilots where there is less harmonic distortion on the signal.

Columns 3 and 4 indicate the voltage estimation results when the full scale capacitance $C^{\star}_p$ is known. 
Clearly the mean bias over all tests is reduced so that it is close to $0 \%$.
In the experiments however, it is difficult to calibrate one device over another due to lack of high resolution multimeters.
Both digital multimeters and the LMCVT may very well have offsets on the order of $1\%$.
This however, does not take away from the fact that in both cases, the maximum variation over all tests is quite small.
In both cases, the mean error is less than $1 \%$ in the higher voltage cases, where the effect of external interferers is minimal.

These results can be compare to the passive calibration method published in \cite{Moghe2014}.
This comparison is not exact since in \cite{Moghe2014}, 5 minute averages are used and the test voltage levels ranging from 5 kV to 25 kV.
Following Row 2 of Table IV ($HV_2$) in \cite{Moghe2014}, the authors report $\mu(e) = 1.72\%$, $r(e) = (0.1,  14.8)\%$.
In comparison, the results in Table \ref{tab:voltage_estimation_table}, Column 3 show single cycle tests where the mean error is $\mu(e) = 0.71 \%$.
Additionally, the maximum errors are (excluding the lowest voltage level $r(e) = (0.03, 1.79)$.  
Comparing the variation of errors, \cite{Moghe2014} reports $\sigma(e) = 4.2\%$, while the active calibration procedure attains $\sigma(e) = 0.05\%$.
Although, the mean error decreases by $60 \%$, the maximum error decreases by $90 \%$  while the standard deviation decreases by $98 \%$.
It should be noted that this comparison is overly conservative since the reported values in Table \ref{tab:voltage_estimation_table} are for single cycle estimates, while \cite{Moghe2014} reports 5 minute averaged values.
%
\subsection{Pilot Frequency Selector} 
\label{subsubsection-Pilot-Frequency-Selector}

Since the line magnitude can be orders of magnitude larger than the injected pilot signal, the presence of harmonics or spurious signal a the exact pilot frequency can be a source of error. 
Care must be taken in deciding what frequency is chosen.  
\begin{figure}[h]
\hspace{-7mm}   
\subfigure[][]{  
\label{fig:good_bad_pilot}
\includegraphics[scale=0.26]{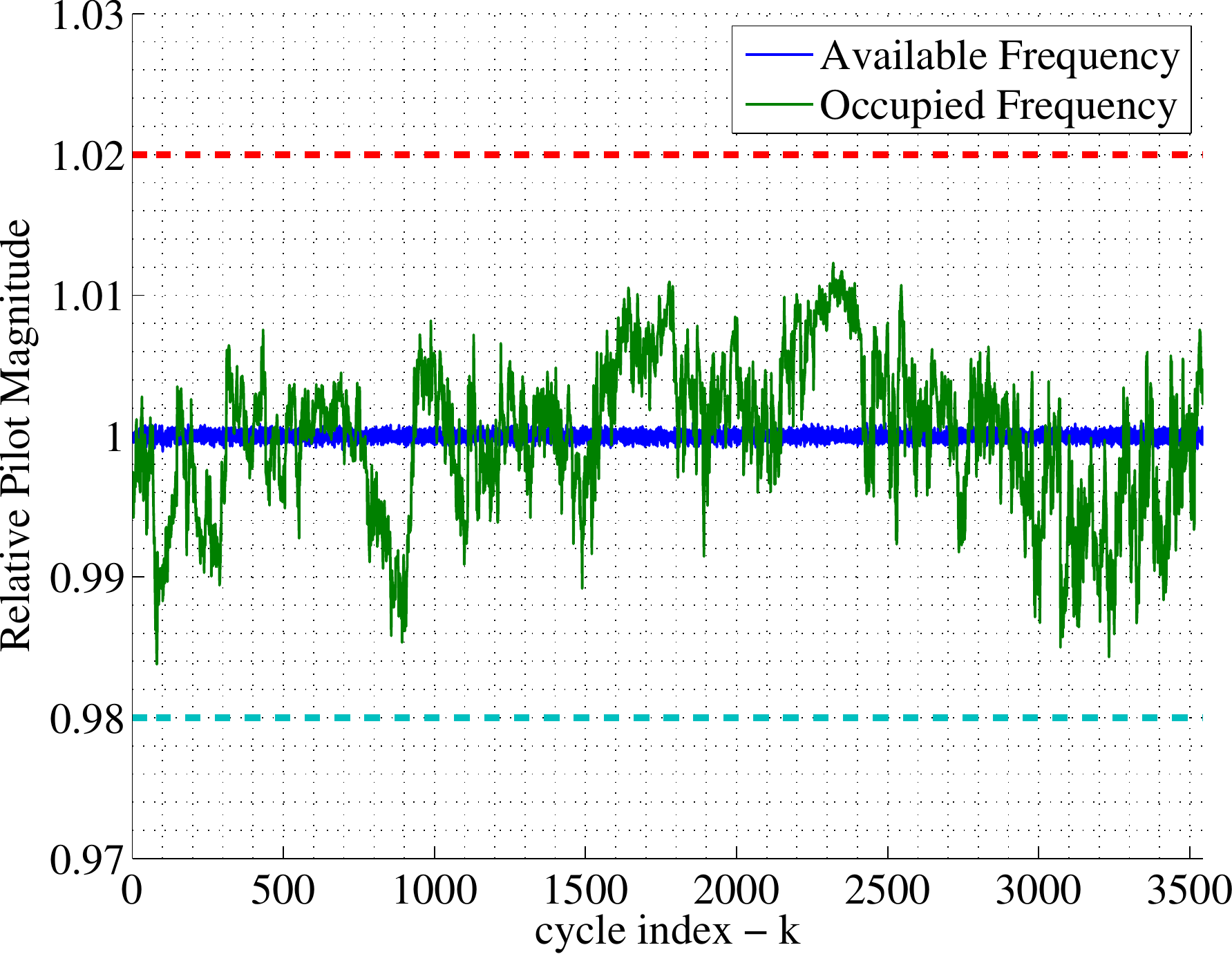} 
}  
\hspace{-4mm}
\subfigure[][]{  
\label{fig:pendulum_test}  
\includegraphics[scale=0.26]{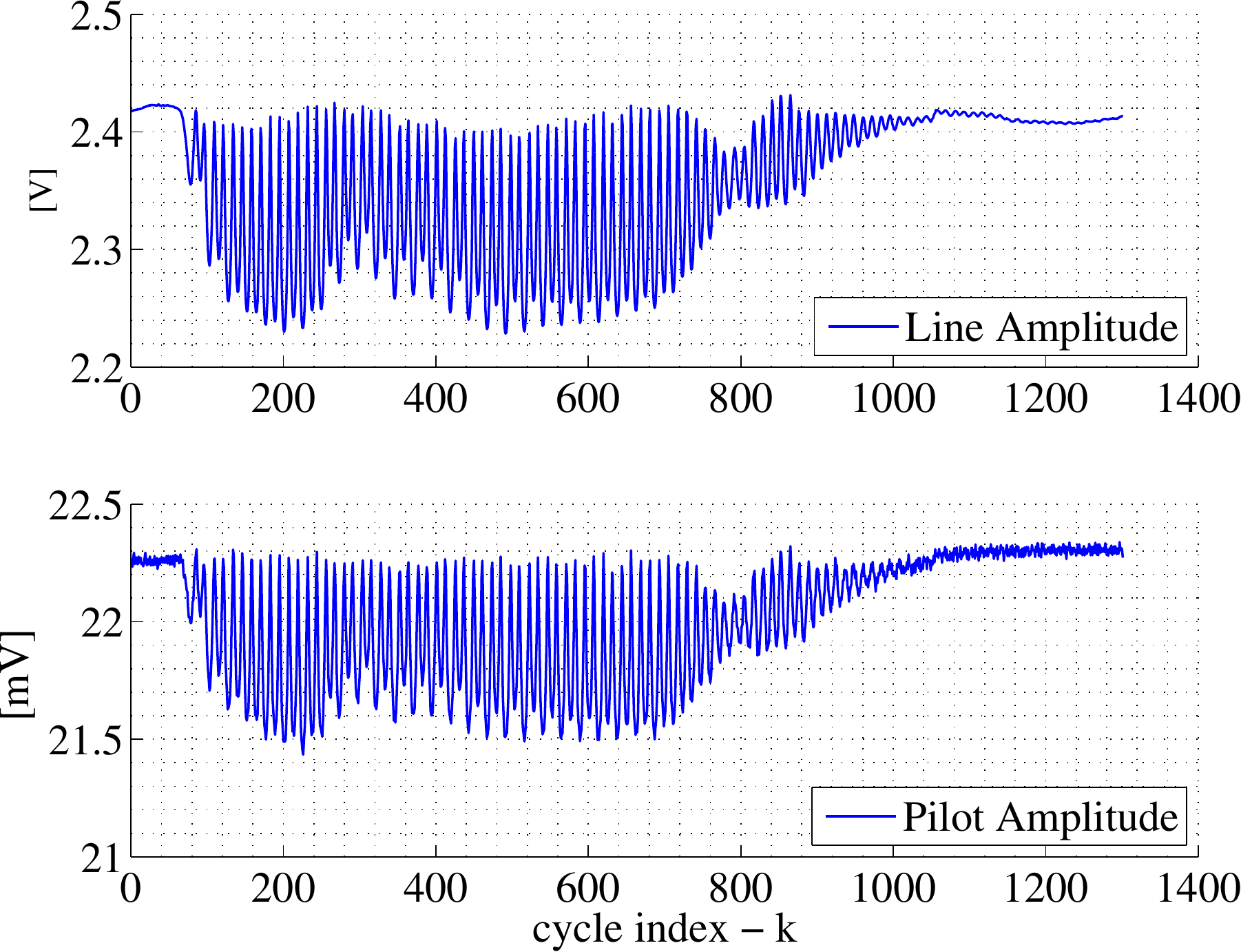} 
}   
\caption{
\ref{fig:good_bad_pilot}       
Experiment showing two frequencies used for pilot magnitude estimation: $f_1 = 3.2$ kHz and $f_2 = 5$ kHz.
Pilot $f_1$ sees an interference signal, while $f_2$ does not.
The result is clear difference in the variation of each pilot amplitude: the clear frequency sees additive low variance white noise, while the occupied frequency experiences high variance autoregressive process.
}
\end{figure}
%
%
An experiment is performed where two non-harmonic pilot frequencies were chosen, 3.2 kHz and 5 kHz, and are used for active calibration.
In this case, the 3.2 kHz frequency contained energy from other sources, while the 5 kHz frequency has none. 
These are referred to as \textit{available} and \textit{occupied} pilot frequencies.
In the reconstruction of the pilot frequency, there is a very clear difference between the two signals shown in Figure \ref{fig:good_bad_pilot}.
In the experiment the least square estimate in \eqref{eq:least_square_sinusoid}, is solved every 1/60 seconds.
Using a sample rate of 55 kHz, this leads to $\sim 917$ samples for estimation, leading to high accuracy in the NLLS procedure.

Observing the two situations indicate a clear difference in each signal type.  
A clear channel frequency leads to constant receive pilot amplitude, assuming that the environment is fixed.
In this case, the error is of very low variance, uncorrelated and gaussian.
On the other hand, when the pilot frequency is occupied, the received amplitude is highly correlated with a variance an order of magnitude larger than in the clear channel case.
There is likely to be confusion between interfering signals at a chosen pilot, and actual environmental changes that can lead to error if a single pilot signal in a fixed frequency is used.
If multiple frequencies, or multiple pilots are used, this error can be averted.
\subsection{Capacitance Change Detection} 
\label{subsubsection-Capacitance-Change-Detection}

If the pilot frequency is clear of any interference sources, $\beta[k]$ can be used to detect changes in the environment.
At low enough signal frequencies it can be assumed that all frequencies will see the same $C_P(t)$.
Any environmental disturbance affecting the 60 Hz line can be distinguished from actual voltage changes since it will be seen on a clear pilot frequency.
An experiment illustrating this was performed (see Appendix for image).
A metallic pendulum was built and used to bring a grounded surface repeatedly close to the shell of the device, as what would happen in some fast changing environmental change on the line.
A line voltage of 1.2 kV was applied on the main line, with a pilot signal of 10 V at 5 kHz.
Again, a 1/60 second capture window was chosen for processing. 
Although this can be varied, it's implausible that changes in the physical environment will occur at a very high rate.
Figure \ref{fig:pendulum_test} shows clearly that disturbances on both the 60 Hz line and 5 kHz pilot signal can be detected due to the sensitivity of the pilot signal.
The frequency of the pendulum given by $\frac{1}{2 \pi} \sqrt{\frac{g}{l}} \approx 1.3 Hz$ matches closely to the frequency of detected oscillation.
The output of the detector will be a bad data quality flag indicating accuracy compromise of the system.
Since environmental disturbances do no constantly occur on a stable transmission lines, the impact of measurement continuity is likely to be minor.

The variance of the pilot estimate in a fixed environment will be crucial to the performance of any change point detector.
There is a tradeoff in computational resources dedicated to more refined pilot tracking; for example, (1) tracking and removing various harmonics, adaptively tracking and estimating their frequencies (3) length and bandwidth of bandpass filters. 

It is clear that the change detection problem and the capacitance tracking problem can be solved separately.
However, provided better modeling of how probe capacitances behave in the real world, further joint optimization is possible but not explored here.

%
%
%
\section{Future Work and Open Problems}
\label{section-future-work-and-open-problems}

In order to further improve the accuracy of such technology and enable full deployment, the following future areas must be explored:

\begin{enumerate}
\item \textit{Interference Decoupling:} 
In the case of multi-conductor systems, there will be multiple coupled sources at similar voltage levels as well as measurements at each source.
Work is needed to design methods and algorithms where multiple phasor measurements are used to decouple each voltage source in this situation.
\item \textit{Bias Correction via multi-Frequency Pilot Mechanism:}
As shown in Section \ref{subsubsection-Capacitance-Estimation}, the offset in received pilot magnitude leads to an offset in the recovered probe capacitance as well as voltage estimate.
A multi-frequency pilot can potentially remove this bias and improve the estimated voltages, and should be explored.
\item \textit{Adaptive Frequency Selection:}
Work on efficient frequency probing/selection algorithms must be done to ensure that the pilot frequencies used can be used for probe estimation.
Figure \ref{fig:good_bad_pilot}, illustrates that simple time series testing can detect such changes properly.
\item \textit{Environmental Change Detection:}
Efficiently detecting environmental changes from the measured pilot signals to guarantee that any change in $\alpha[k]$ is in fact from line voltage change.
\end{enumerate}

\section{Conclusions and Future Work}     
\label{section-conclusion}

This work presents a method of achieving high accuracy line mountable capacitive voltage transducers for high voltage applications.
The method relies on computing the probe-to-ground capacitance in real time using out of band pilot signal injection.
The concept is tested in experiments and a roadmap of a functional prototype is given.

\begin{appendix}
%
%
%
\section{Testing Environment}
\subsection{Low Voltage Test Environment}
\label{subsection-Low-Voltage-Test-Environment}

The test environment we use is a metallic mesh cage connected to ground.
Low voltage tests are performed are susceptible to 60 Hz interference at 120/240 V which is an order of magnitude larger than the line voltage.
Therefore a metallic cage is useful in making an interference free environment. 
Consider prototyping a system with $V_L = 10$ V.
Unfortunately, all nearby interference sources are scattered randomly in a real laboratory environment with $V_I = 120$ V.

\subsection{High Voltage Test Environment}
\label{subsection-High-Voltage-Test-Environment}

\begin{figure}[h]
\centering

\subfigure[][]{
\includegraphics[scale=0.30]{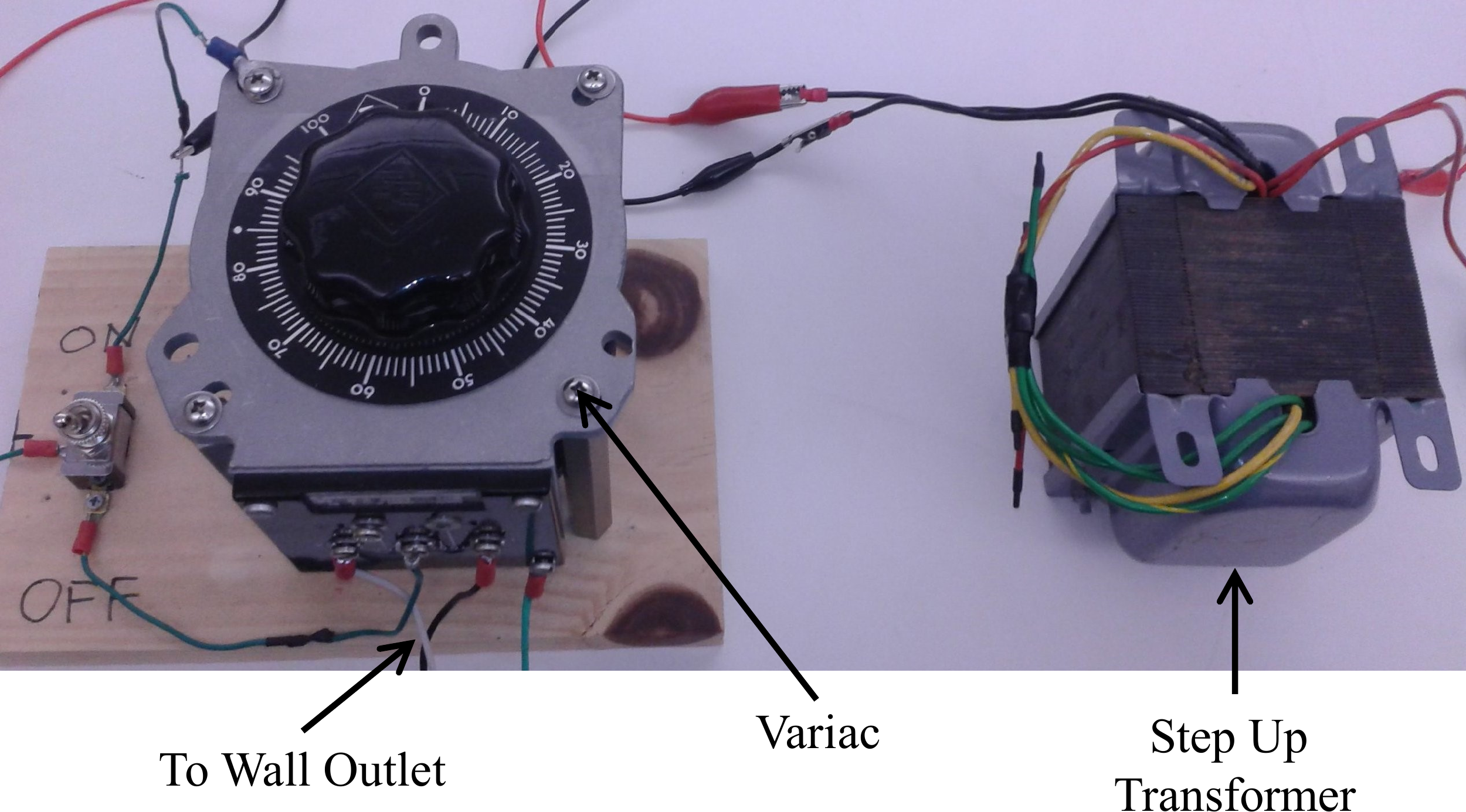}  
\label{fig:VL2_generation}
}
\caption{High voltage test setup used to generate a maximum of 1.2 kV in laboratory setting.}
\end{figure}

Figure \ref{fig:VL2_generation} shows the test setup used to generate a high voltage AC source.
A variac connected to a step up transformer can generate a moderated AC signal for full scale voltage testing.
\section{Experimental Evaluation Physical Model of Body Capacitive Voltage Sensor}
\subsection{Verifying Probe Capacitance in Low Voltage Testing}
\label{subsection-Verifying-Probe-Capacitance-in-Low-Voltage-Testing}
\begin{figure}[h]
\centering

\subfigure[][]{
\includegraphics[scale=0.30]{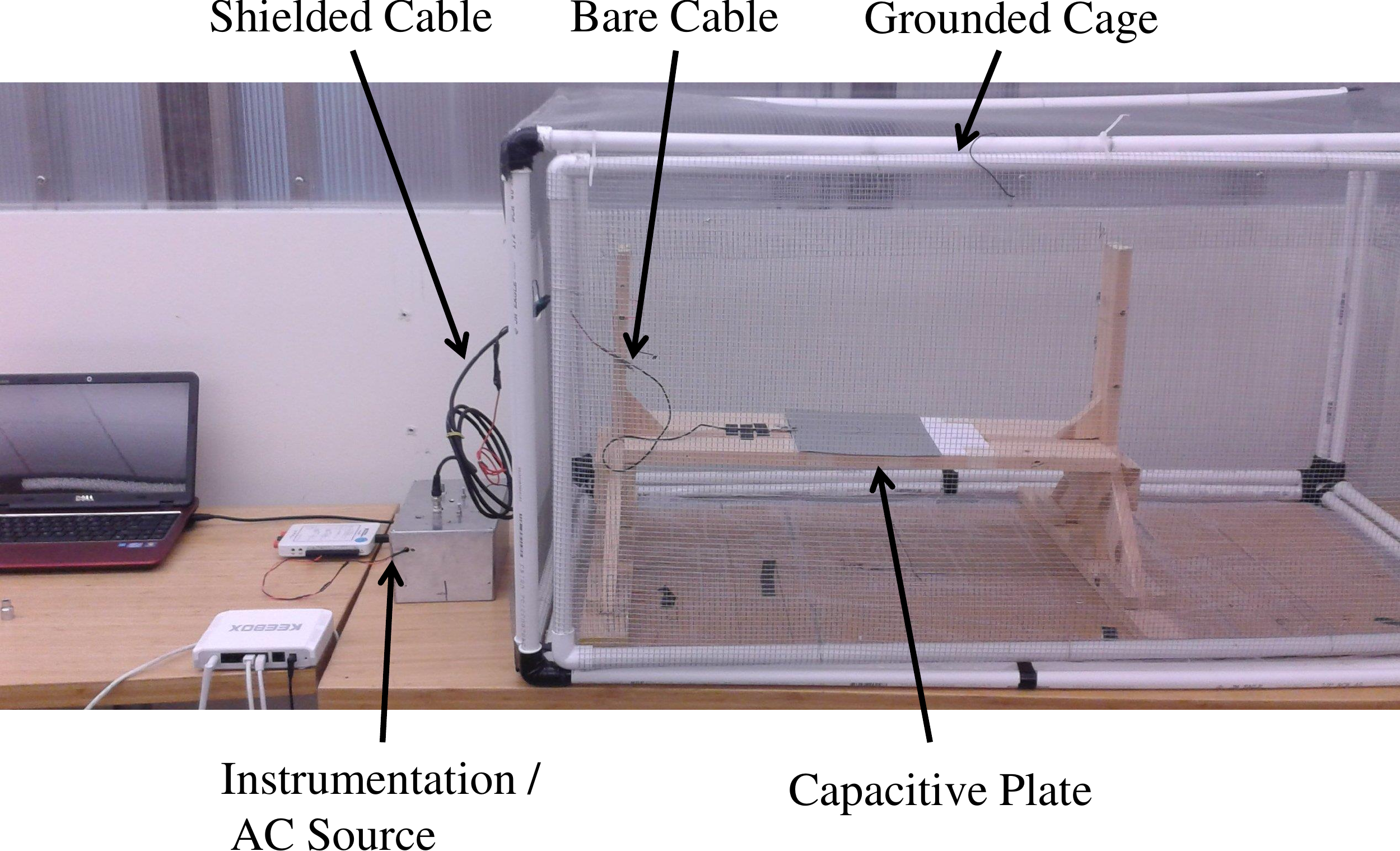}  
\label{fig:plate_test_setup}
}
\caption{\ref{fig:plate_test_setup} Experiment setup for evaluating the voltage input/output relation and body capacitance $C_P$ of an ideal rectangular plate in fixed environment.}
\end{figure}

\begin{table}[h]
\centering
\caption{Ideal Plate Capacitance Evaluation.  Simulation (S) and Experimental (E) values are in close agreement.} 
\label{tab:ideal-plate-test}
\begin{tabular}{@{}cccccc@{}}
\toprule
Plate $\#$  &  & Dim. [in x in]& $C_P$ [pF] (S)  &$C_P$ [pF] (E) \\
\cmidrule{1-1}                                \cmidrule{3-5}
1                &  & 12x12 &   15.00    &  14.99 \\
2                &  & 8x8     &  10.34     &  10.27 \\
3                &  & 4x4     &    4.78     &  4.88   \\
4                &  & 2x2     &    2.3       &  1.84   \\
5                &  & 1x1     &    1.2       &  0.38   \\
\bottomrule
\end{tabular}
\end{table}

Here we demonstrate the basic body capacitive effect. 
First given a set of sheet metal plates of various sizes, we compute the theoretical capacitance of the probe with no charged device in proximity of the of the charged plate.
This is done in COMSOL Multiphysics simulation environment using the electrostatic simulation package. 

\begin{figure}[h]
\centering
\subfigure[][]{
\label{fig:plate_test_vin_vout}
\includegraphics[scale=0.23]{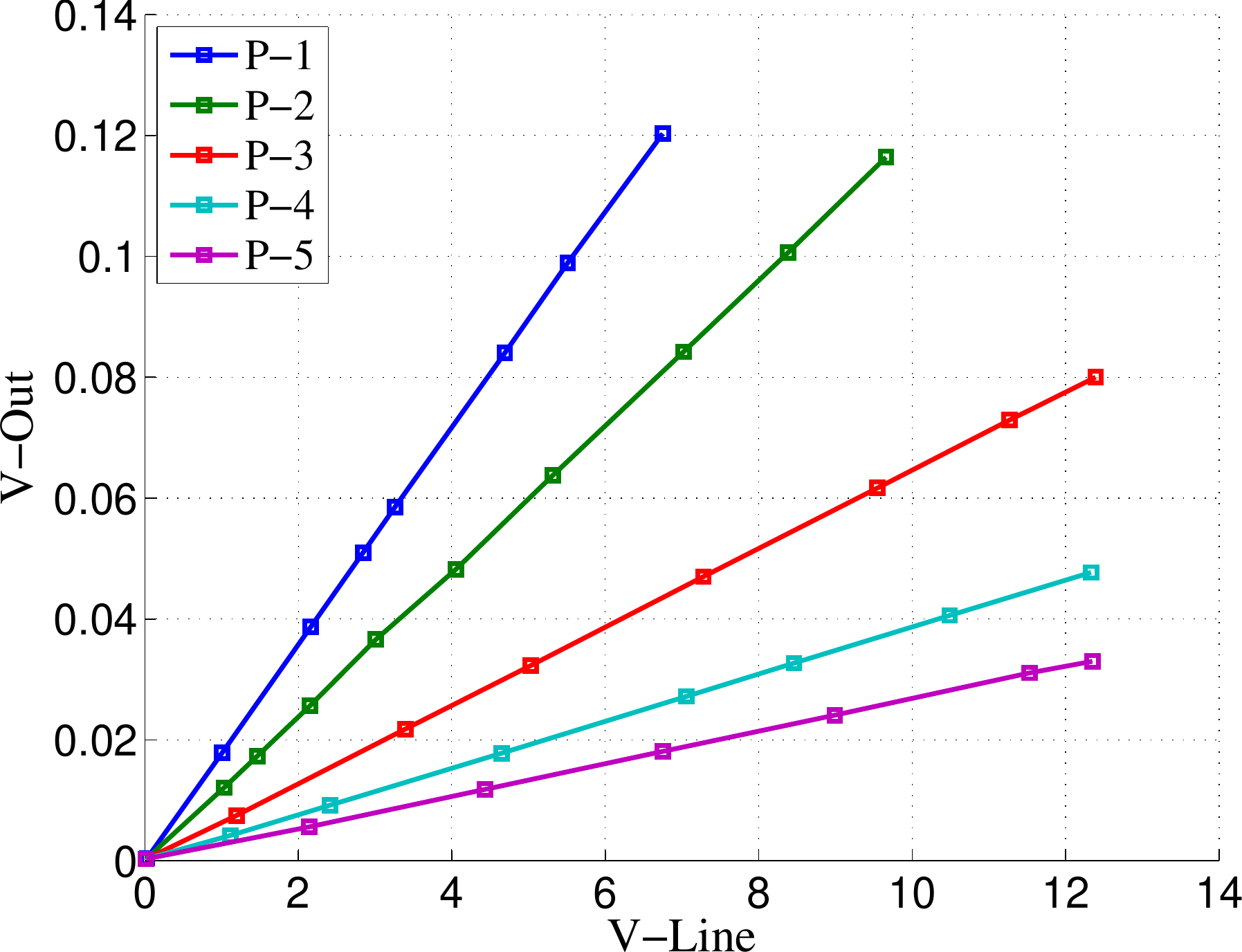}  
}
\subfigure[][]{
\label{fig:proximity_test_result}
\includegraphics[scale=0.23]{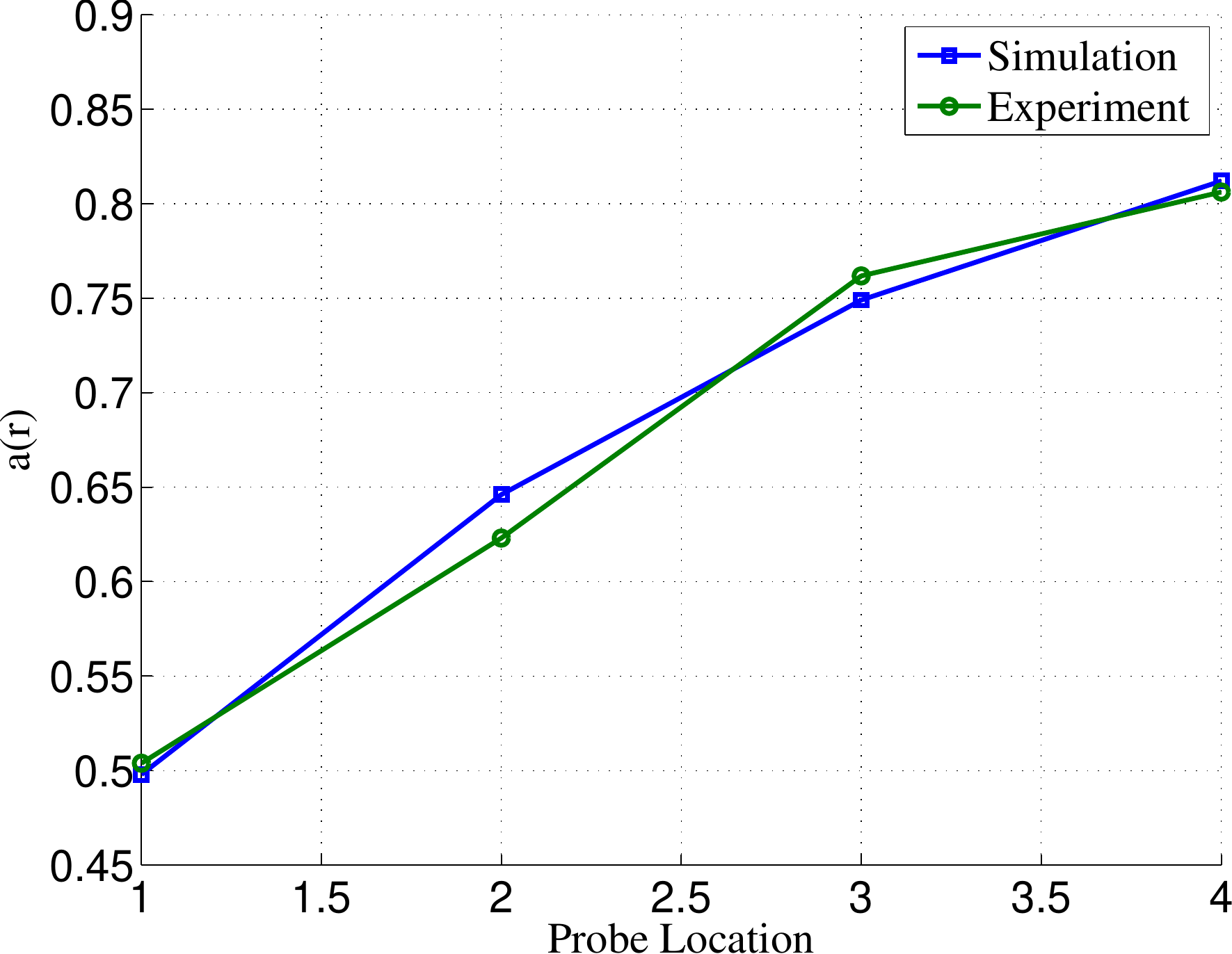}  
}
\caption{ 
\ref{fig:plate_test_vin_vout}
Input/Output voltage relationship of body capacitive probe.  
Results from plate test verifying the theoretical linear relationship from Eq. \eqref{eq:ideal-cap-input-output-with-interference} .
\ref{fig:proximity_test_result}
Simulation and experimental values of $\alpha(r)$ at various distances from the charged power line.  
Results show the values are close leading to acceptance of the model.
}
\end{figure}

The test arrangement of this is shown in Figure \ref{fig:plate_test_setup}.
To experimentally evaluate the charged plate with an ideal voltage source with no proximity effect reducing the effective capacitance we perform the following.

\begin{enumerate}
\item Place the voltage source (NI Hardware) and charge sensing device (metallic box) outside the grounded cage.
\item Connect the plate to a shielded cable which becomes unshielded at the entrance of the cage
\item Compute the total capacitance of the cable and plate body capacitance.  
This is done by measuring the rms input/output voltage (Figure \ref{fig:plate_test_vin_vout}). 
This leads to an experimental value of $C_P$.
\item Repeat step 3 with the cable alone.  The difference in capacitances being that of the plate alone.
\end{enumerate}  

Table \ref{tab:ideal-plate-test} shows that for larger plates, the simulated value matches very closely to the electrostatic simulation.
However the simulation and experiment diverge for smaller plates.  
Regardless, the experiments validate the basic premise of the body capacitor model.
Figure \ref{fig:plate_test_vin_vout} shows a linear relation between $V_{L}$ and $V_{CAP}$ for each plate, validating the model.
   
\subsection{Effective Capacitance}
\label{subsection-effective-capacitance}

\begin{figure}[h]
\centering
\subfigure[][]{
\label{fig:proximity_test_setup}
\includegraphics[scale=0.45]{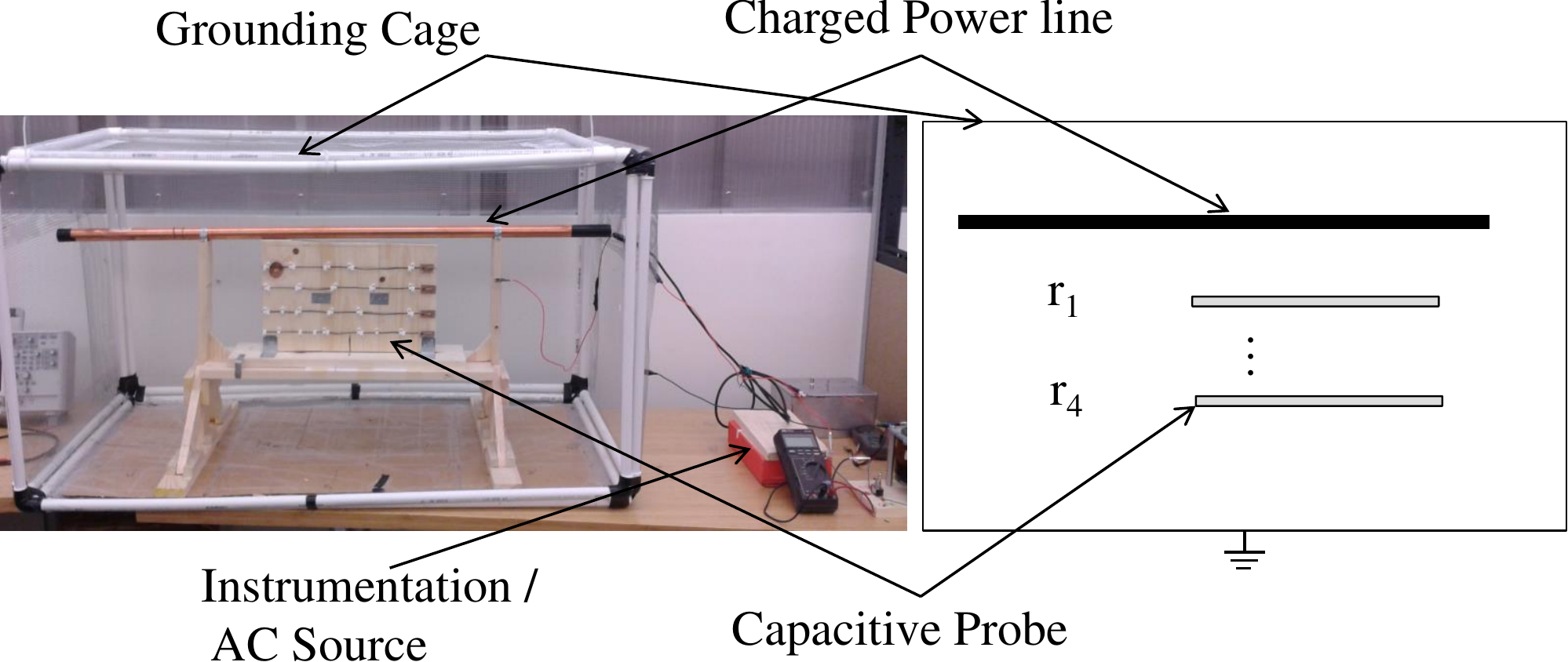}  
}
\subfigure[][]{
\label{fig:cable_iso_surface}
\includegraphics[scale=0.35]{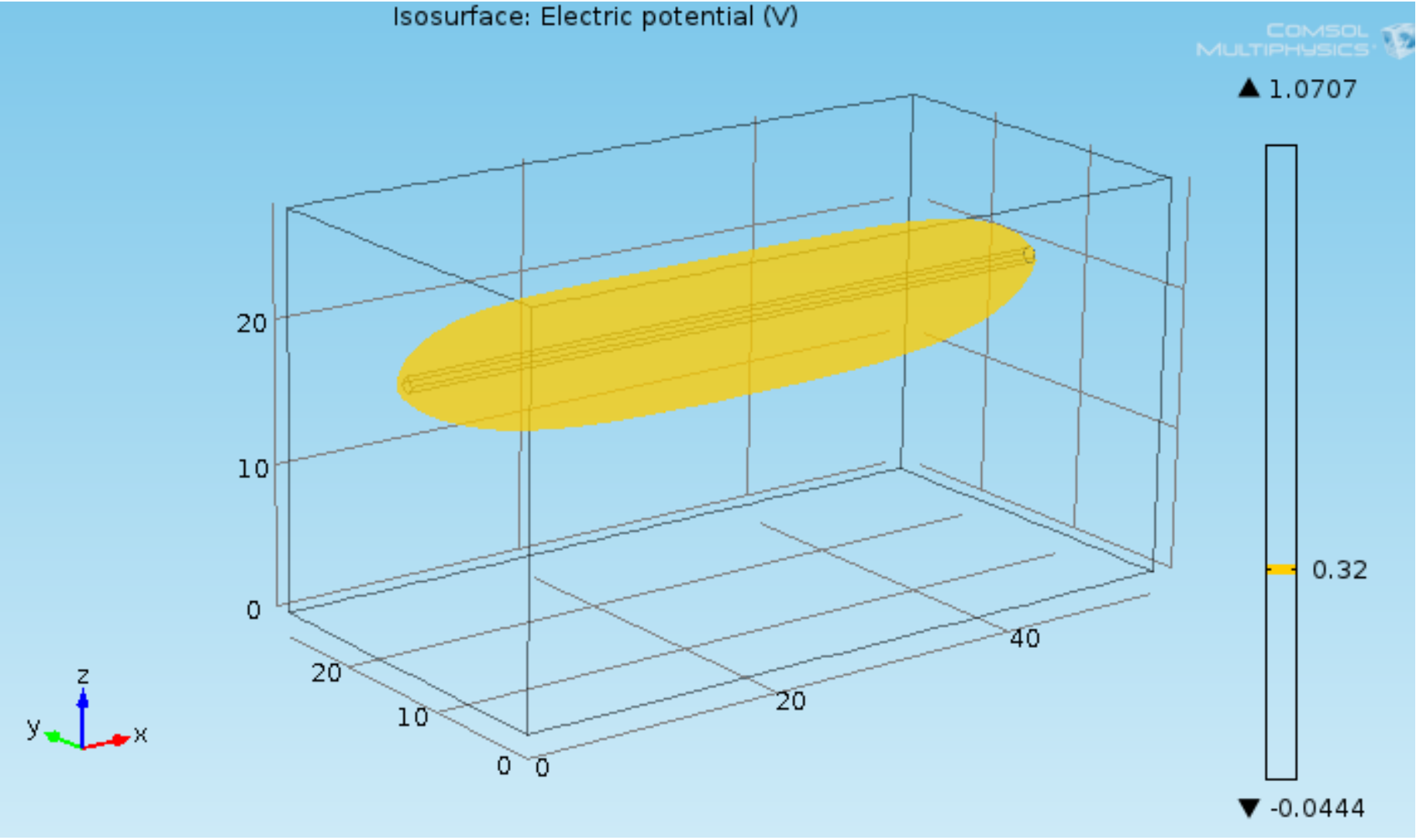}  
}
\caption{   
\ref{fig:proximity_test_setup} 
Experiment setup for evaluating the proximity effect on the probe capacitance.  
Setup similar to the plate test, except with wire probe and a charged conducted (at $V_L$) to simulate a power line the device will be connected to. 
\ref{fig:cable_iso_surface} 
Iso-surface of voltage level computed in COMSOL for specific geometry.
}
\end{figure}

We show a method of verifying the proposed model for the effect of power line proximity to the body capacitance of a probe.
Figure \ref{fig:proximity_test_setup} shows a test setup inside the grounded cage.
Like the test in Section \ref{subsection-Verifying-Probe-Capacitance-in-Low-Voltage-Testing} the instrumentation is placed outside of the cage to prevent any unintentional charge to act on the probe.
However, there is an external field from the copper power line in the experiment.
The probes are copper wires set on at fixed distances from the power line.  
This is done to minimize measurement error in evaluating various distances.

A similar procedure is conducted as in Section \ref{subsection-Verifying-Probe-Capacitance-in-Low-Voltage-Testing} where the capacitance of the cable and the probe is computed first;
then the capacitance of the cable alone.
The difference of the two being the experimental effective capacitance under a charged ($C_p$) and uncharged ($C_P$) power line.
An experimental value of $\alpha(r) = 1 - \frac{C_p}{C_P}$ is computed, where $C_p$ and $C_P$ are computed at each distance.
A theoretical $\alpha(r_k)$ can be computed easily without computing the individual $C_p$ or $C_P$ since $\alpha(r) = \frac{V(r)}{V_L}$.
Given the simulation tool, we can evaluate $V(r)$ from the geometry as shown in Figure \ref{fig:cable_iso_surface}.
For simplicity we evaluate the CFD solution for the point voltage along the same position of the cable probe.
An average of the solution vector is then $V(r)$. 

Figure \ref{fig:proximity_test_result} shows the results of measuring the quantity $\alpha(r)$ at the various locations of capacitive probe.
The results show considerable agreement between the two models thus leading us to accept the model.
The proximity effect is important to consider in the design of a final body capacitive probe. 
Notice that in location 4 which is only 4.3 cm from the line, $\alpha(r)$ is 0.8. 
Any probe that is closer to the power line will have an effective capacitance that is much smaller than in the uncharged case. 
Therefore, there is a minimum separation we need between the probe and the power line so as to have a value of $C_p$ that works in practice.

\section{Probe Capacitance Disturbance Experiment Setup}
\label{subsection-effective-capacitance}

\begin{figure}[h]
\centering   
\subfigure[][]{
\label{fig:pendulum_test_setup}
\includegraphics[scale=0.3]{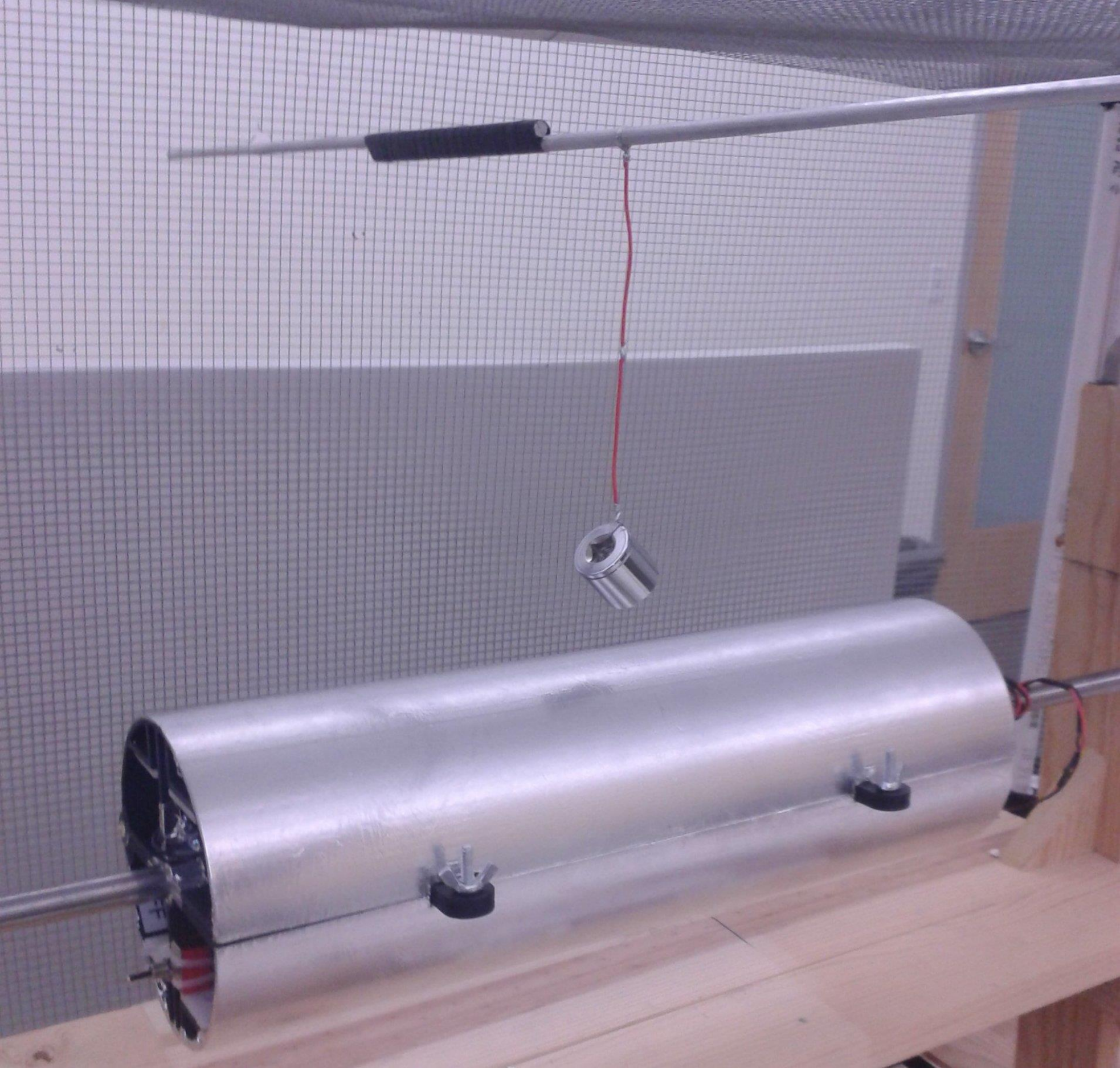}  
}
\caption{ 
Conductor pendulum placed in setup near charged probe. 
Pendulum excited by hand to illustrate sensitivity of pilot signal to external disturbances.
}
\end{figure}

Figure \ref{fig:pendulum_test_setup} shows the pendulum test setup to generate a periodic disturbance that is easily detectable by the pilot signaling mechanism.

\begin{table}[!htb]
\centering     
\caption{Signal Processing Architecture: Variable Definitions}   
\label{tab:signal_processing_architecture}   
\begin{tabular}{@{}lll@{}}   
\toprule        
 Variable                                 & Definition                                               & Typical Value \\
\cmidrule{1-1}                         \cmidrule{2-2}                                           	\cmidrule{3-3} 
$n$             	               	    & ADC sample Index 			            		&  \\  
$F_{ADC}$                            & ADC Frame Rate 			            		&  50 KS/s\\
$F_{frame}$                          & Device Output Rate 		                    &  60 S/s\\                  
$k$             	               	    & Device output frame index   	           		& \\
$v[n]$                         	    & Differential Input sample 	                   &       \\ 
$v_L[n], v_B[n]$                   & Low Pass/Band Pass Filter of $v[n]$      &\\
$\alpha[k], \phi[k]$                & Mains Magnitude/Phase Estimate          & \\ 
$M$                                      & Number of pilot signals 			   	     & 1 \\
$f_{p, m}, V_{p,m}$              & Pilot Signal Frequency/Magnitude          &  3,5 KHz~/~10 V\\
$\beta_m[k]$                        & Recovered Pilot magnitude                     &  \\
$\hat{C}_p[k]$                      & Probe Capacitance Estimate   	           & 16 pF\\
$n_{PPS}$                           & GPS Pulse Per Second Output   	           & \\
$\hat{V}_L[k], \hat{\phi}[k]$  & Voltage Phasor Estimate      		  	& \\
\bottomrule 
\end{tabular}
\end{table}
   
\end{appendix}

\bibliographystyle{unsrt}
\bibliography{tps_bib}

\end{document}